\documentclass[letterpaper,twocolumn,english,prd,nofootinbib,
preprintnumbers,showpacs,showkeys]{revtex4}
\usepackage[T1]{fontenc}
\usepackage[latin1]{inputenc}
\usepackage{babel}
\usepackage{color}
\usepackage{graphics}
\usepackage{subfigure}

\usepackage{mathrsfs} 
\renewcommand{\vec}[1]{{\bf #1}}

\begin{document}

\newcommand{\wh}[1]{\widehat{#1 }}

\newcommand{\wt}[1]{\widetilde{#1 }}

\newcommand{\xhat}{\widehat{x}}

\newcommand{\zhat}{\widehat{z}}

\newcommand{\qTh}{\wh{q}_{T}}

\newcommand{\shat}{\widehat{s}}

\newcommand{\that}{\widehat{t}}

\newcommand{\uhat}{\widehat{u}}

\newcommand{\ov}[1]{\overline{#1 }}

\newcommand{\ASud}{{\cal A}}

\newcommand{\BSud}{{\cal B}}

\newcommand{\sw}{\sigma _{\wt{W}}}

\newcommand{\sfo}{\sigma _{FO}}

\newcommand{\sasy}{\sigma _{ASY}}

\newcommand{\stot}{\sigma _{TOT}}

\newcommand{\qt}{q_{T}}

\newcommand{\scrP}{\mathscr {P}}

\newcommand{\scrH}{{\mathscr H}}

\preprint{hep-ph/0210082, Cavendish-HEP-02/15, CTEQ-209, 
MSUHEP-20930}

\title{Resummation of transverse momentum and mass logarithms\\
in DIS heavy-quark production}

\author{
P. M. Nadolsky$ ^{1} $, 
N. Kidonakis$ ^{2} $, 
F. I. Olness$ ^{1} $,
C.-P. Yuan$ ^{3} $}

\affiliation{
$ ^{1} $Department of Physics, Southern Methodist University, Dallas,
TX 75275-0175, U.S.A.\\
$ ^{2} $Cavendish Laboratory, University of Cambridge, Madingley Road,
Cambridge CB3 0HE, United Kingdom\\
$ ^{3} $Department of Physics \& Astronomy, Michigan State University,
East Lansing, MI 48824, U.S.A.}

\date{14th October 2002}

\begin{abstract}
Differential distributions for heavy quark production depend on the heavy
quark mass and other momentum scales, which can yield additional 
large logarithms
and inhibit accurate predictions. Logarithms involving the heavy quark mass
can be summed in heavy quark parton distribution functions in the ACOT 
factorization
scheme. A second class of logarithms involving the heavy-quark transverse
momentum can be summed using an extension of Collins-Soper-Sterman (CSS)
formalism. We perform a systematic summation of logarithms of both types,
thereby obtaining an accurate description of heavy-quark differential 
distributions
at all energies. Our method essentially combines the ACOT and CSS approaches.
As an example, we present angular distributions for bottom quarks produced
in parity-conserving events at large momentum transfers $ Q $ at the $ ep $
collider HERA.
\end{abstract}

\pacs{12.38 Cy, 13.60 -r}
\keywords{deep inelastic scattering, 
summation of perturbation theory, heavy flavor production}
\maketitle

\section{Introduction\label{sec:Intro}}

In recent years, significant attention was dedicated to exploring properties
of heavy-flavor hadrons produced in lepton-nucleon deep inelastic scattering
(DIS). On the experimental side, the Hadron-Electron Ring Accelerator (HERA)
at DESY has generated a large amount of data on the production of charmed
\cite{Adloff:1996xq, Breitweg:1997mj, Adloff:1998vb, Breitweg:1999ad, Adloff:2001zj}
and bottom mesons \cite{Adloff:1999nr, Breitweg:2000nz, BottomDISH1, BottomDISZEUS1, BottomZEUS2}.
At present energies (of order 300 GeV in the $ ep $ center-of-mass frame),
a substantial charm production cross section is observed in a wide range
of Bjorken $ x $ and photon virtualities $ Q^{2} $, and charm quarks
contribute up to 30\% to the DIS structure functions. 

On the theory side, Perturbative Quantum Chromodynamics (PQCD) provides a
natural framework for the description of heavy-flavor production. Due to
the large masses $ M $ of the charm and bottom quarks 
($ M^{2}\gg \Lambda _{QCD}^{2} $),
the renormalization scale can be always chosen in a region where the effective
QCD coupling $ \alpha _{S} $ is small. Despite the smallness of
 $ \alpha _{S} $,
perturbative calculations in the presence of heavy flavors are not without
intricacies. In particular, care in the choice of a factorization scheme
is essential for the efficient separation of the short- and long-distance
contributions to the heavy-quark cross section. This choice depends on the
value of $ Q $ as compared to the heavy quark mass $ M $. The key issue
here is whether, for a given renormalization and factorization scale 
$ \mu _{F}\sim Q $,
the heavy quarks of the $ N $-th flavor are treated as \emph{partons}
in the incoming proton, \emph{i.e.}, whether one calculates the QCD 
beta-function
using $ N $ active quark flavors and introduces a parton distribution
function (PDF) for the $ N $-th flavor. A related, but separate, issue
is whether the mass of the heavy quark can be neglected in the hard cross
section without ruining the accuracy of the calculation. 

Currently, several factorization schemes are available that provide different
approaches to the treatment of these issues. Among the mass-retaining 
factorization
schemes, we would like to single out the fixed flavor number factorization
scheme (FFN scheme), which includes the heavy-quark contributions exclusively
in the hard cross section 
\cite{Gluck:1982cp,Gluck:1988uk,Nason:1989zy,Laenen:1992cc,Laenen:1993zk,Laenen:1993xs};
and massive variable flavor number schemes (VFN schemes), which introduce
the PDFs for the heavy quarks and change the number of active flavors by
one unit when a heavy quark threshold is crossed 
\cite{Collins:1998rz,Aivazis:1994pi,Kniehl:1995em,Buza:1998wv,Thorne:1998ga,Thorne:1998uu,Cacciari:1998it,Chuvakin:1999nx,Kramer:2000hn}.
Further details on these schemes can be found later in the paper. Here we
would like to point out that, were the calculation done to all orders of
$ \alpha _{S} $, the FFN and massive VFN schemes would be exactly equivalent.
However, in a finite-order calculation the perturbative series in one scheme
may converge faster than that in the other scheme. In particular, the FFN
scheme presents the most economic way to organize the perturbative calculation
near the heavy quark threshold, \emph{i.e.}, when $ Q^{2}\approx M^{2} $.
At the same time, it becomes inappropriate at $ Q^{2}\gg M^{2} $ due to
powers of large logarithms $ \ln \left( Q^{2}/M^{2}\right)  $ in the hard
cross section. In the VFN schemes, these logarithms are summed through all
orders in the heavy-quark PDF with the help of the 
Dokshitzer-Gribov-Lipatov-Altarelli-Parisi
(DGLAP) equation \cite{Dokshitzer:1977sg,Gribov:1972ri,Altarelli:1977zs};
hence the perturbative convergence in the high-energy limit is preserved.
In their turn, the VFN schemes may converge slower at $ Q^{2}\approx M^{2} $,
mostly because of the violation of energy conservation in the heavy-quark
PDF's in that region. Recently an optimal VFN scheme was proposed that 
compensates
for this effect \cite{Tung:2001mv}. 

In this paper, we would like to concentrate on the analysis of semi-inclusive
differential distributions (\emph{i.e.}, distributions depending on additional
kinematical variables besides $ x $ and $ Q $). We will argue that
finite-order calculations in any factorization scheme do not satisfactorily
describe such distributions due to additional large logarithms besides the
logarithms $ \ln (Q^{2}/M^{2}) $. To obtain stable predictions, all-order
summation of these extra logarithmic terms is necessary. 

\begin{figure}
{\centering \resizebox*{1\columnwidth}{!}{
\includegraphics{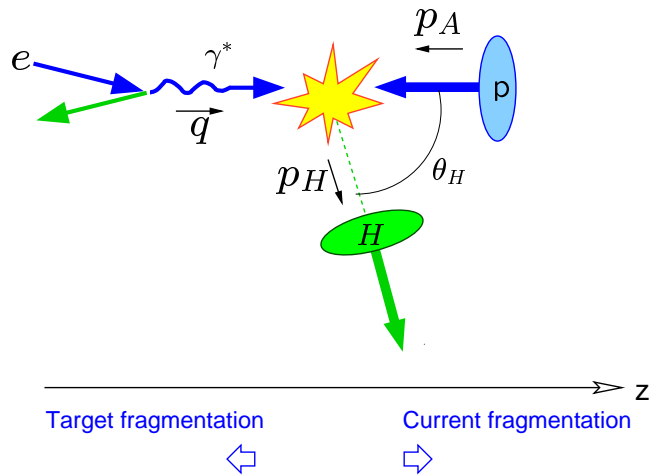}} \par}
\caption{\label{fig:GammaP}The parity-conserving semi-inclusive 
production \protect$ e+p\rightarrow e+H+X\protect $
of heavy hadrons in the \protect$ \gamma ^{*}p\protect $ c.m.~reference
frame. The resummation effects considered here are important in the current
fragmentation region \protect$ \theta _{H}\rightarrow 0\protect $, \emph{i.e.},
when the final-state heavy quark \protect$ h\protect $ closely follows
the direction of its escape in the 
\protect$ {\cal O}(\alpha _{S}^{0})\protect $
flavor-excitation process \protect$ \gamma ^{*}+h\rightarrow h\protect $.}
\end{figure}

The extra logarithms are of the form 
$ (\alpha _{S}^{n}/q_{T}^{2})\ln ^{m}(q_{T}^{2}/Q^{2}) $,
$ 0\leq m\leq 2n-1 $, where $ q_{T}=p_{T}/z $ denotes the transverse
momentum $ p_{T} $ of the heavy hadron in the $ \gamma ^{*}p $ center-of-mass
(c.m.)~reference frame rescaled by the variable 
$ z\equiv (p_{A}\cdot p_{H})/(p_{A}\cdot q) $.
Here $ p_{A}^{\mu },\, q^{\mu }, $ and $ p_{H}^{\mu } $ are the momenta
of the initial-state proton, virtual photon, and heavy hadron, respectively.
Our definitions for the $ \gamma ^{*}p $ c.m.~frame and hadron momenta
are illustrated by Fig.~\ref{fig:GammaP}. The resummation of these logarithms
is needed when the final-state hadron escapes in the current fragmentation
region (\emph{i.e.}, close the direction of the virtual photon in the
$ \gamma ^{*}p $
c.m. frame, where the rate is the largest). In the current fragmentation
region, the ratio $ q_{T}^{2}/Q^{2} $ is small; therefore, the terms
 $ \ln ^{m}(q_{T}^{2}/Q^{2}) $
compensate for the smallness of $ \alpha _{S} $ at each order 
of the perturbative
expansion. If hadronic masses are neglected, such logarithms can be summed
through all orders in the impact parameter space resummation formalism 
\cite{Collins:1993kk, Meng:1996yn, Nadolsky:1999kb, Nadolsky:2000ky},
which was originally introduced to describe angular correlations in 
$ e^{+}e^{-} $
hadroproduction \cite{Collins:1981uk, Collins:1982va} and transverse momentum
distributions in the Drell-Yan process \cite{Collins:1985kg}.\footnote{%
The similarity between the multiple parton radiation in semi-inclusive DIS
and the other two processes was known for a long time; see, for instance,
an early paper \cite{Dokshitzer:1978dr}.
} Here the impact parameter $ b $ is conjugate to $ q_{T} $. The results
of Refs.~\cite{Collins:1993kk, Meng:1996yn, Nadolsky:1999kb, Nadolsky:2000ky}
are immediately valid for semi-inclusive DIS (SIDIS) production of light
hadrons ($ \pi ,K,... $) at $ Q $ of a few GeV or higher, and for 
semi-inclusive
heavy quark production at $ Q^{2}\gg M^{2} $. To describe heavy-flavor
production at $ Q^{2}\sim M^{2} $, the massless 
$ q_{T} $-resummation
formalism must be extended to include the dependence on the heavy-quark mass
$ M $.

In this paper, we perform such extension in the Aivazis-Collins-Olness-Tung
(ACOT) massive VFN scheme \textcolor{black}{\cite{Aivazis:1994pi} with the
optimized treatment of the} threshold region \cite{Tung:2001mv}. We adopt
a {}``bottom-up{}'' approach to the development of such mass-dependent
resummation.\footnote{%
An alternative {}``top-down{}'' approach will require the analysis of leading
regions in the high-energy limit and derivation of the evolution equations
that retain terms with positive powers of $ M/Q $. Such analysis could
involve methods similar to those discussed in Ref. \cite{Collins:1999dz}.
} We start by separately reviewing the massive VFN scheme in the inclusive
DIS and $ q_{T} $ resummation in the massless SIDIS. We then discuss a
combination of these two frameworks in a joint resummation of the logarithms
$ \ln (Q^{2}/M^{2}) $ and $ \ln (q_{T}^{2}/Q^{2} $). \textcolor{black}{}As
a result, we obtain a unified description of fully differential heavy-hadron
distributions at all $ Q^{2} $ above the heavy quark threshold. It is
well known that the finite-order calculation does not satisfactorily treat
the current fragmentation region for any choice of the factorization scheme.
In contrast, the proposed massive extension of the $ q_{T} $-resummation
accurately describes the current fragmentation region in the whole range
$ Q^{2}>M^{2} $.

The present study is interesting for two phenomenological reasons. Firstly,
the quality of the differential data will improve greatly within the next
few years. By 2006, the upgraded collider HERA will accumulate an integrated
luminosity of $ 1\mbox {\, fb}^{-1} $ \cite{Greenshaw:2002wu}, \emph{i.e.},
more than eight times the final integrated luminosity from its previous runs.
Studies of the heavy quarks in DIS are also envisioned at the 
proposed high-luminosity
Electron Ion Collider \cite{EICWhiteBook} and THERA \cite{Abramowicz:2001qt}.
Eventually these experiments will present detailed distributions both at
small ($ Q^{2}\approx M^{2} $) and large ($ Q^{2}\gg M^{2} $) momentum
transfers.

Secondly, the knowledge of the differential distributions is essential for
the accurate reconstruction of inclusive observables, such as the charm component
of the structure function $ F_{2}(x,Q^{2}) $. At HERA, 
\textcolor{black}{40-60\%
of the charm production events occur outside the detector acceptance region,
notably at small transverse momenta of the heavy quarks. To determine 
$ F_{2}^{c}(x,Q^{2}) $,
those events should be reconstructed with the help of some theoretical model,
which so far was the $ {\cal O}(\alpha _{S}^{2}) $ calculation in the
FFN scheme 
\cite{Laenen:1993zk,Laenen:1993xs,Harris:1995pr,Harris:1995tu,Harris:1998zq}
incorporated in a massless parton showering generator. As mentioned above,
for inclusive observables the FFN scheme works the best not far from the
threshold and becomes unstable at $ Q^{2}\gg M^{2} $, where the VFN scheme
is more appropriate. In more detail, the VFN scheme describes 
$ F_{2}^{c}(x,Q^{2}) $
better than the FFN scheme when $ Q^{2} $ exceeds 
$ 20\, (\mbox {GeV}/c)^{2} $,}
\textcolor{black}{\emph{i.e.}}\textcolor{black}{, 
roughly when $ Q^{2}/M^{2}>10 $
\cite{Buza:1998wv}. The transition to the VFN scheme occurs faster at smaller
$ x $, where the c.m. energy of the $ \gamma ^{*}p $ collision is much
larger than $ M $. For bottom quark production, the estimate 
$ Q^{2}/M^{2}\gtrsim 10 $
corresponds to $ Q^{2}\gtrsim 200\, (\mbox {GeV}/c)^{2} $. The VFN calculation
can also be extended down to the mass threshold to uniformly describe the
whole range $ Q^{2}>M^{2} $. Since the proposed resummation calculation
is formulated in the VFN scheme, it provides a better alternative to the
finite-order calculation in the FFN scheme due to its correct treatment of
differential distributions at all values of $ Q^{2} $. }

As an example, we apply the developed method to the leading-order 
flavor-creation
and flavor-excitation processes in the production of bottom mesons at HERA.
We find that the resummed cross section for this process can be described
purely by means of perturbation theory due to the large mass of the bottom
quark. Our predictions can be tested in the next few years once the integrated
luminosity at HERA approaches $ 1\mbox {\, fb}^{-1}. $ Essentially the
same method can be applied to charm production. In that case, however, the
resummed cross section is sensitive to the nonperturbative large-$ b $
contributions due to the smaller mass of the charm quarks, and the analysis
is more involved. Since the goal of this paper is to discuss the basic principles
of the massive $ q_{T} $-resummation, we leave the study of the charm
production and other phenomenological aspects for future publications.

The paper has the following structure. Section \ref{sec:ACOT} reviews the
application of the ACOT factorization scheme \cite{Aivazis:1994pi} and its
simplified version \cite{Collins:1998rz, Kramer:2000hn} to the description
of the inclusive DIS structure functions. Section \ref{sec:CSS} recaps basic
features of the $ b $-space resummation formalism \emph{}in \emph{massless}
SIDIS. Section \ref{sec:MassiveCSS} discusses modifications in the resummed
cross section to incorporate the dependence on the heavy-quark mass $ M $.
In Section~\ref{sec:PhotonGluon}, we present a detailed calculation of
the mass-dependent resummed cross sections in the leading-order flavor-creation
and flavor-excitation processes. Section \ref{sec:NumericalResults} presents
numerical results for polar angle distributions in the production of bottom
quarks at HERA. Appendix~\ref{Appendix:Chg} contains details on the calculation
of the $ {\cal O}(\alpha _{S}) $ mass-dependent part of the resummed cross
section. In Appendix~\ref{Appendix:FO}, we present explicit expressions
for the $ {\cal O}(\alpha _{S}) $ finite-order contributions from the
photon-gluon channel. Finally, Appendix~\ref{Appendix:KinematicalCorrection}
discusses in detail the optimization of the ACOT scheme when it is applied
to the differential distributions in the vicinity of the threshold region.

\section{Overview of the factorization scheme \label{sec:ACOT}}

\subsection{Factorization in the presence of heavy quarks}

In this Section, we discuss the application of the Aivazis-Collins-Olness-Tung
(ACOT) factorization scheme \cite{Aivazis:1994pi} to inclusive DIS observables,
for which this scheme yields accurate predictions both at asymptotically
high energies and near the heavy-quark threshold.

In the inclusive DIS, the factorization in the presence of heavy flavors
is established by a factorization theorem \cite{Collins:1998rz}, which we
review under a simplifying assumption that only one heavy flavor $ h $
with the mass $ M $ is present. Let $ A $ denote the incident hadron.
According to the theorem, the contribution $ F_{h/A}(x,Q^{2}) $ of $ h $
to a DIS structure function $ F(x,Q^{2}) $ (where $ F(x,Q^{2}) $ is
one of the functions $ F_{1}(x,Q^{2}) $ or $ F_{2}(x,Q^{2}) $ 
in parity-conserving
DIS) can be written as
\begin{eqnarray}
 &  & F_{h/A}(x,Q^{2})=\sum _{a}\int _{\chi _{a}}^{1}\frac{d\xi }{\xi }C_{h/a}\left( \frac{\chi _{a}}{\xi },\frac{\mu _{F}}{Q},\frac{M}{Q}\right) \nonumber \\
 &  & \times f_{a/A}\left( \xi ,\, \left\{ \frac{\mu _{F}}{m_{q}}\right\} \right) +{\mathcal{O}}\left( \frac{\Lambda _{QCD}}{Q}\right) .\label{F} 
\end{eqnarray}
 Here the summation over the internal index $ a $ includes both light
partons (gluons $ G $ and light quarks), as well as the heavy
quark $ h $. This representation is accurate up to the non-factorizable
terms that do not depend on $ M $ and can be ignored when $ Q\gg \Lambda _{QCD} $.
The non-vanishing term on the r.h.s. is written as a convolution integral
of parton distribution functions $ f_{a/A}\left( \xi ,\, \{\mu _{F}/m_{q}\}\right)  $
and coefficient functions $ C_{h/a}(\chi _{a}/\xi ,\, \mu _{F}/Q,\, M/Q) $.
The convolution is realized over the hadron light-cone momentum fraction
$ \xi  $ carried by the parton $ a $. The coefficient function depends
on the flavor-dependent {}``scaling variable{}'' $ \chi _{a} $ discussed
below. The parton distributions and coefficient functions are separated by
an \emph{arbitrary} factorization scale $ \mu _{F} $ such that $ f_{a/A} $
depends only on $ \mu _{F} $ and quark masses $ \{m_{q}\}\equiv m_{u},m_{d},...,M $;
and $ C_{h/a} $ depends only on $ \mu _{F},\, M, $ and $ Q $. As
a result of this separation, all logarithmic terms $ \alpha _{S}^{n}\ln ^{k}\left( \mu _{F}/m_{q}\right)  $
\emph{with light-quark masses} are included in the PDF's, where they are
summed through all orders using the DGLAP equation. Note that in the massless
approximation such logarithms appear in the guise of collinear poles $ 1/\epsilon ^{k} $
in the procedure of dimensional regularization. The logarithms $ \ln ^{k}(\mu _{F}/M) $
with \emph{the heavy-quark mass} $ M $ are included either in $ C_{h/a} $
or $ f_{a/A} $ depending on the factorization scheme in use. 

In the reference frame where the momentum of the incident hadron $ A $
in the light-cone coordinates is 
\begin{displaymath}
p_{A}^{\mu }=
\left\{ p_{A}^{+},\frac{m_{A}^{2}}{2p_{A}^{+}},\vec{0}_{T}\right\} ,
\end{displaymath}
(where $ p^{\pm }\equiv \left( p^{0}\pm p^{3}\right) /\sqrt{2} $), the
quark PDF can be defined in terms of the quark field operators $ \psi_q (x) $
as \cite{Collins:1982uw}
\begin{eqnarray}
 &  & f_{q/A}\left( \xi ,\left\{ \frac{\mu _{F}}{m_{q}}\right\} \right) =\overline{\sum _{spin}}\, \overline{\sum _{color}}\int \frac{dy^{-}}{2\pi }e^{-i\xi p_{A}^{+}y^{-}}\nonumber \\
 &  & \times \langle p_{A}|\bar{\psi_q }(0,y^{-},\vec{0}_{T})\nonumber \\
 &  & \times {\cal P}\exp \left\{ -ig\int _{0}^{y^{-}}dz^{-}{\mathscr A}^{+}(0,z^{-},\vec{0}_{T})\right\} \nonumber \\
 &  & \times \frac{\gamma ^{+}}{2}\psi_q (0)|p_{A}\rangle .\label{f} 
\end{eqnarray}
Here $ {\cal P}\exp {\{...\}} $ is the path-ordered exponential of the
gluon field $ {\mathscr A}_{\nu }(x) $ in the gauge $ \eta \cdot {\mathscr A}=0 $.
The r.h.s. is averaged over the spin and color of $ A $ and summed over
the spin and color of $ q $. A similar definition exists for the gluon
PDF. The dependence of $ f_{a/A}(\xi ,\{\mu _{F}/m_{q}\}) $ on $ \mu _{F} $
is induced in the process of renormalization of ultraviolet (UV) singularities
that appear in the bilocal operator on the r.h.s. of Eq.~(\ref{f}). In
general, the PDF is a nonperturbative object; however, it can be calculated
in PQCD when $ \mu _{F}\gg \Lambda _{QCD} $, and the incident hadron $ A $
is replaced by a parton. This feature opens the door for the calculation
of $ F_{h/A}(x,Q^{2}) $ for any hadron $ A $ through the conventional
sequence of calculating $ C_{h/a}(\chi _{a}/\xi ,\mu _{F}/Q,M/Q) $ in
parton-level DIS and convolving it with the phenomenological parameterization
of the nonperturbative PDF $ f_{a/A}(\xi ,\{\mu _{F}/m_{q}\}) $. In the
inclusive DIS, it is convenient to choose $ \mu _{F}\sim Q $ to avoid
the appearance of the large logarithm $ \ln (\mu _{F}/Q) $ in $ C_{h/a}(\chi _{a}/\xi ,\mu _{F}/Q,M/Q) $.

The factorized representation (\ref{F}) is valid in all factorization schemes.
The specific factorization scheme is determined by (a) the procedure for
the renormalization of the UV singularities; and (b) the prescription for
keeping or discarding terms with positive powers of $ M/Q $ in the coefficient
function $ C_{h/a} $. The choice (a) determines if the logarithms $ \ln ^{k}(\mu _{F}/M) $
are resummed in the heavy-flavor PDF or not. With respect to each of two
issues, the choice can be done independently. For instance, the $ \overline{MS} $
factorization scheme uses the dimensional regularization to handle the UV
singularities, but does not uniquely determine the choice (b). Hence, it
is not necessary in this scheme to always neglect $ M $ in the coefficient
function and expose the heavy-quark mass singularities as poles in the dimensional
regularization.

The ACOT scheme belongs to the class of the variable flavor number (VFN)
factorization schemes~\cite{Collins:1978wz} that change the renormalization
prescription when $ \mu _{F} $ crosses a threshold value $ \mu _{thr} $.
It is convenient to choose $ \mu _{thr} $ for the flavor $ h $ to be
equal to $ M $, since the logarithms $ \ln ^{k}\left( \mu _{F}/M\right)  $
vanish at that point. If $ \mu _{F}<M $, all graphs with internal heavy-quark
lines are renormalized by zero-momentum subtraction. If $ \mu _{F}>M $,
these graphs are renormalized in the $ \overline{MS} $ scheme. The masses
of the light quarks are neglected everywhere, and graphs with only light
parton lines are always renormalized in the $ \overline{MS} $ scheme. 

The physical picture behind the ACOT prescription is simple: the heavy quark
is excluded as a constituent of the hadron for sufficiently low energy (an
$ N-1 $ flavor subscheme), but the heavy quark is included as a constituent
for sufficiently high energies (an $ N $ flavor subscheme). The renormalization
by zero-momentum subtraction below the threshold leads to the explicit decoupling
of the heavy-quark contributions from light parton lines. As one consequence
of the decoupling, all perturbative components of the heavy-quark PDF vanish
at $ \mu _{F}<M $, so that a nonzero heavy-quark PDF may appear only through
nonperturbative channels, such as the {}``intrinsic heavy quark mechanism{}''
\cite{Brodsky:1980pb}. Since the size of such nonperturbative contributions
remains uncertain, they are not considered in this study. At $ \mu _{F}>M $, a
non-zero heavy-quark PDF $ f_{h/A} $ is introduced, which is evolved together
with the rest of the PDFs with the help of the mass-independent $ \overline{MS} $
splitting kernels. The initial condition for $ f_{h/A}(\xi ,\mu _{F}) $
is obtained by matching the factorization subschemes at $ \mu _{F}=M $.
At order $ \alpha _{S} $, this condition is trivial:
\begin{equation}
f_{h/A}(\xi ,\mu _{F}=M)=0.
\end{equation}
At higher orders, the initial value of $ f_{h/A}(\xi ,\mu _{F}) $ is given
by a superposition of light-flavor PDF's \cite{Buza:1998wv}. A simple illustration
of these issues is given in Appendix~\ref{Appendix:Chg}. 

\begin{figure}
{\centering \resizebox*{!}{3cm}{\includegraphics{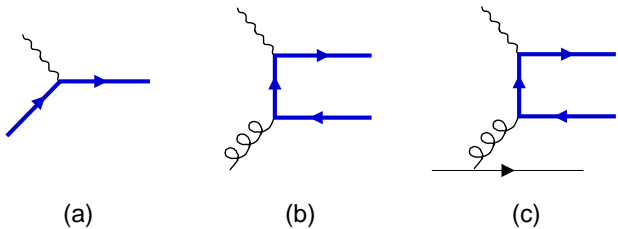}} \par}

\caption{\label{fig:diag2} Basic subprocesses in the ACOT scheme: (a) flavor excitation
\protect$ \gamma ^{*}+h\rightarrow h\protect $ at \protect$ {\mathcal{O}}(\alpha _{S}^{0})\protect $;
(b) gluon flavor creation (photon-gluon fusion) \protect$ \gamma ^{*}+G\rightarrow h+\bar{h}\protect $
at \protect$ {\mathcal{O}}(\alpha _{S}^{1})\protect $; (c) light-quark
flavor creation \protect$ \gamma ^{*}+q\rightarrow (\gamma ^{*}+G)+q\rightarrow (h+\bar{h})+q\protect $
at \protect$ {\cal O}(\alpha _{S}^{2})\protect $. The thick and thin solid
lines correspond to the heavy quark \protect$ h\protect $ and light quarks
\protect$ q=u,d,s\protect $, respectively.}
\end{figure}

The ACOT scheme possesses another important property: the coefficient function
$ C_{h/a} $ in this scheme has a finite limit as $ Q\rightarrow \infty  $,
which coincides with the expression for the coefficient function obtained
in the massless $ \overline{MS} $ scheme with $ N $ active flavors.
This happens because the mass-dependent terms in $ C_{h/a} $ contain only
positive powers of $ M/Q $, while the quasi-collinear logarithms $ \ln (\mu _{F}/M) $
are resummed in $ f_{h/A}(\xi ,\mu _{F}) $. As a consequence of the introduction
of $ f_{h/A}(\xi ,\mu _{F}) $, the coefficient function $ C_{h/a} $
includes subprocesses of three classes: 

\begin{itemize}
\item \emph{flavor excitation}, where the parton $ a $ is a heavy quark; 
\item gluon \emph{flavor creation}, where $ a $ is a gluon; 
\item and light-quark \emph{flavor creation}, where $ a $ is a light quark.
\end{itemize}
In contrast, in the FFN scheme \cite{Gluck:1982cp, Gluck:1988uk, Nason:1989zy, Laenen:1992cc, Laenen:1993xs, Laenen:1993zk}
only the flavor-creation processes are present. The lowest-order diagrams
for each class are shown in Fig.~\ref{fig:diag2}. 

The subsequent parts of the paper consider the processes shown in Figs.~\ref{fig:diag2}a
and \ref{fig:diag2}b. Note that we count the order of diagrams according
to the explicit power of $ \alpha _{S} $ in the coefficient function,
\emph{i.e.}, $ {\cal O}(\alpha _{S}^{0}) $ in Fig.~\ref{fig:diag2}a,
$ {\cal O}(\alpha _{S}^{1}) $ in Fig.~\ref{fig:diag2}b, and $ {\cal O}(\alpha _{S}^{2}) $
in Fig.~\ref{fig:diag2}c. This counting does not apply to the whole structure
function $ F_{h/A}(x,Q^{2}) $ in Eq.~(\ref{F}) when the heavy-quark
PDF is itself suppressed by $ \alpha _{S}/\pi  $ near the mass threshold
\cite{Olness:1988ep,Barnett:1988jw,Sullivan:2001ry}. In that region, an
$ {\cal O}(\alpha _{S}^{n}) $ \emph{flavor-excitation} contribution has
roughly the same order of magnitude as the $ {\cal O}(\alpha _{S}^{n+1}) $
\emph{flavor-creation} contribution. We return to this issue in the discussion
of numerical results in Section~\ref{sec:NumericalResults}, where we interpret
the combination of the $ {\cal O}(\alpha _{S}^{0}) $ \emph{flavor-excitation}
contribution (Fig.~\ref{fig:diag2}a) and $ {\cal O}(\alpha _{S}^{1}) $
\emph{flavor-creation} contribution (Fig.~\ref{fig:diag2}b) as a first
approximation at $ Q\approx M $.

\subsection{Simplified ACOT Formalism}

Of several available versions of the ACOT scheme, our calculation utilizes
its modification advocated by Collins~\cite{Collins:1998rz}, which we identify
as the Simplified ACOT (S-ACOT) formalism \cite{Kramer:2000hn}. It has the
advantage of being easy to state and of allowing relatively simple calculations.
This simplicity could be crucial for implementing the massive VFN prescription
at the next-to-leading order in the global analysis of parton distributions.

In brief, this prescription is stated as follows.

\begin{quote}
\textit{Simplified ACOT (S-ACOT) prescription}: Set $ M $ to zero in the
calculation of the coefficient functions $ C_{h/a} $ for the incoming
heavy quarks: that is, 
\begin{displaymath}
C_{h/h}\left( \frac{\chi _{h}}{\xi },\frac{\mu _{F}}{Q},\frac{M}{Q}\right) \rightarrow C_{h/h}\left( \frac{\chi _{h}}{\xi },\frac{\mu _{F}}{Q},0\right) .
\end{displaymath}

\end{quote}
It is important to note that this prescription is not an approximation; it
correctly accounts for the full mass dependence \cite{Collins:1998rz}. It
also tremendously reduces the complexity of flavor-excitation structure functions,
as they are given by the light-quark result. In the specific case considered
here, the heavy quark mass in the S-ACOT scheme should be retained only in
the $ \gamma ^{*}+G\rightarrow h+\bar{h} $ subprocess (Fig.~\ref{fig:diag2}b).
Another important consequence will be discussed in Section~\ref{sec:MassiveCSS},
where we show that the S-ACOT scheme leads to a simpler generalization of
the $ q_{T} $-resummation to the mass-dependent case.

\subsection{\label{sec:DISrescalingVariable}The scaling variable }

Finally, we address the issue of the most appropriate variables $ \chi _{a} $
($ a=G,u,d,s,... $) in the convolution integral (\ref{F}). In a massless
calculation, $ \chi _{a} $ are just equal to Bjorken $ x $, since all
momentum fractions $ \xi  $ between $ x $ and unity are allowed by
energy conservation. This simple relation does not hold in the massive case.
For instance, in the charged-current heavy quark production $ W^{\pm }+q\rightarrow h, $
where $ h $ is present in the final, but not the initial, state, a simple
kinematical argument leads to the conclusion that the longitudinal variable
in the flavor-excitation processes should be rescaled by a mass-dependent
factor, as $ \chi _{h}=x\left( 1+M^{2}/Q^{2}\right)  $ \cite{Barnett:1988jw}. 

In the flavor-excitation subprocesses of the neutral-current heavy quark
production (\emph{e.g.}, $ \gamma ^{*}+h\rightarrow h $), typically no
rescaling correction was made. The presence of a heavy quark in \textit{both}
the initial and final states of the hard scattering suggested that no kinematical
shift was necessary, \textit{i.e.}, $ \chi _{h}=x $. This assumption has
been recently questioned by a new analysis~\cite{Tung:2001mv}. Specifically,
Tung \textit{et al.}~note that the heavy quarks in the hadron come predominantly
from gluons splitting into quark-antiquark pairs. Hence the heavy quark $ h $
initiating the hard process must be accompanied by the unobserved $ \bar{h} $
in the beam remnant. When both $ h $ and $ \bar{h} $ are present, the
hadron's light-cone momentum fraction carried by the incoming parton cannot
be smaller than $ x\left( 1+4M^{2}/Q^{2}\right)  $, which is larger than
the minimal momentum fraction $ \xi _{min}=x $ allowed by the single-particle
inclusive kinematics. The factor of $ 4M^{2} $ arises from the threshold
condition for $ h $ and $ \bar{h} $. This effect can be accounted for
by evaluating the flavor-excitation cross sections at the scaling variable
$ \chi _{h}=x\left( 1+4M^{2}/Q^{2}\right)  $. 

In brief, the rule proposed in Ref.~\cite{Tung:2001mv} is to use $ \chi _{a}=x\left( 1+4M^{2}/Q^{2}\right)  $
in flavor-excitation processes (Fig.~\ref{fig:diag2}a) and $ \chi _{a}=x $
in flavor-creation processes (Figs.~\ref{fig:diag2}b and \ref{fig:diag2}c)
when calculating inclusive cross sections. \textcolor{black}{However, to
correctly describe the differential distributions of the final-state hadron,
we have to generalize the above rule for semi-inclusive observables. This
generalization is discussed in Appendix~\ref{Appendix:KinematicalCorrection},
where the proper scaling variable for fully differential finite-order cross
sections is found to be $ \chi _{h}=x\left( 1+M^{2}/\left( z(1-z)Q^{2}\right) \right)  $
(cf.~Eq.~(\ref{chi21})). }

\section{Massless transverse momentum resummation\label{sec:CSS}}

We now turn to the differential distributions of the heavy-flavor cross sections.
Specifically, we consider the production of a heavy-quark hadron $ H $
via the process $ e(\ell )+A(p_{A})\rightarrow H(p_{H})+e(\ell' )+X. $
This reaction is illustrated in Fig.~\ref{fig:GammaP} for the specific
case when $ A $ is a proton. In much of the discussion, we will find it
convenient to amputate the external lepton legs and work with the photon-hadron
process $ \gamma ^{*}(q)+A(p_{A})\rightarrow H(p_{H})+X $ in the photon-hadron
c.m.~frame. Given the conventional DIS variables $ Q^{2}=-q^{2} $ and
$ x=Q^{2}/(2p_{A}\cdot q) $, as well as the Lorentz invariant $ S_{eA}\equiv (\ell +p_{A})^{2} $,
we decompose the electron-level cross section into a sum over the functions
$ A_{\rho }(\psi ,\varphi ) $ of the lepton azimuthal angle $ \varphi  $
and boost parameter $ \psi \equiv \cosh ^{-1}\left( 2xS_{eA}Q^{-2}-1\right)  $
\cite{Meng:1992da,Nadolsky:1999kb}:
\begin{eqnarray}
\frac{d\sigma (e+A\rightarrow e+H+X)}{dxdQ^{2}d\vec{p}_{H}} & \propto  & \sum _{\rho }V_{\rho }(q,p_{A},\vec{p}_{H})\nonumber \\
 & \times  & A_{\rho }(\psi ,\varphi ).
\end{eqnarray}
 This procedure is nothing else but the decomposition over the virtual photon's
helicities \cite{Cheng:1971mx,Lam:1978pu,Olness:1987mv}; hence it is completely
analogous to the tensor decomposition familiar from the inclusive DIS. As
a result of this procedure, the dependence on the kinematics of the final-state
lepton is factorized into the functions $ A_{\rho }(\psi ,\varphi ), $
while the hadronic dynamics affects only the functions $ V_{\rho } $.
In parity-conserving SIDIS, the only contributing angular functions are
\begin{eqnarray}
A_{1}(\psi ,\varphi ) & = & 1+\cosh ^{2}\psi ,\nonumber \\
A_{2}(\psi ,\varphi ) & = & -2,\nonumber \\
A_{3}(\psi ,\varphi ) & = & -\cos \varphi \sinh 2\psi ,\nonumber \\
A_{4}(\psi ,\varphi ) & = & \cos 2\varphi \sinh ^{2}\psi .\label{As} 
\end{eqnarray}

In Section~\ref{sec:ACOT} we found that the ACOT prescription resums logarithms
of the form $ \ln (M^{2}/Q^{2}) $. For the inclusive observables, this
procedure provides accurate predictions throughout the full range of $ x $
and $ Q^{2} $. More differential observables may contain additional large
logarithms in the high-energy limit. In particular, we already mentioned
the logarithms of the type $ (q_{T}^{-2})\alpha _{S}^{n}\ln ^{m}(q_{T}^{2}/Q^{2}),\, 0\leq m\leq 2n-1 $,
which appear when the polar angle $ \theta _{H} $ of the heavy hadron
$ H $ in the $ \gamma ^{*}A $ c.m. frame becomes small (cf.~Fig.~\ref{fig:GammaP}).
Here we chose the $ z $-axis to be directed along the momentum $ \vec{q} $
of the virtual photon $ \gamma ^{*} $. When $ M^{2}\ll Q^{2} $, the
scale $ q_{T} $ is related to $ \theta _{H} $ as
\begin{equation}
q_{T}^{2}=Q^{2}\left( \frac{1}{x}-1\right) \frac{1-\cos \theta _{H}}{1+\cos \theta _{H}};
\end{equation}
hence
\begin{equation}
\lim _{\theta _{H}\rightarrow 0}q_{T}^{2}=Q^{2}\left( \frac{1}{x}-1\right) \left( \frac{\theta _{H}^{2}}{4}+...\right) \rightarrow 0.
\end{equation}

The resummation of these logarithms of soft and collinear origin can be realized
in the formalism by Collins, Soper, and Sterman (CSS) \cite{Collins:1981uk, Collins:1982va, Collins:1985kg, Collins:1989PQCD}.
The result can be expressed as a factorization theorem, which states that
in the limit $ Q^{2}\gg q_{T}^{2},\{m^{2}_{q}\},\Lambda ^{2}_{QCD} $ the
cross section is 
\begin{eqnarray}
 &  & \left. \frac{d\sigma (e+A\rightarrow e+H+X)}{d\Phi }\right| _{q_{T}^{2}\ll Q^{2}}=\frac{\sigma _{0}F_{l}}{2S_{eA}}A_{1}(\psi ,\varphi )\nonumber \\
 &  & \times \int \frac{d^{2}\vec{b}}{(2\pi )^{2}}e^{i\vec{q}_{T}\cdot \vec{b}}\widetilde{W}_{HA}(b,Q,x,z)\nonumber \\
 &  & +{\mathcal{O}}\left( \frac{q_{T}}{Q},\left\{ \frac{m_{q}}{Q}\right\} ,\frac{\Lambda _{QCD}}{Q}\right) .\label{Wmassless} 
\end{eqnarray}
 In this equation, $ b $ is the impact parameter (conjugate to $ \qt  $),
$ d\Phi \equiv dxdQ^{2}dzdq_{T}^{2}d\varphi  $, $ z\equiv (p_{A}\cdot p_{H})/(p_{A}\cdot q) $,
and $ \sigma _{0} $ and $ F_{l} $ are constant factors \textit{}\textit{\emph{given
in}} Eq.~(\ref{sigma0Fl}). As before, $ \{m_{q}\} $ collectively denotes
all quark masses, $ \{m_{q}\}\equiv m_{u},m_{d},...,M. $ At large $ Q^{2} $,
the $ b $-space integral in Eq.~(\ref{Wmassless}) is dominated by contributions
from the region $ b^{2}\lesssim 1/Q^{2} $. In this region, the hadronic
form factor $ \widetilde{W}_{HA}(b,Q,x,z) $ can be factorized in a combination
of parton distribution functions $ f_{a/A}(\xi _{a},\mu _{F}) $, fragmentation
functions $ D_{H/b}(\xi _{b},\mu _{F}) $, and the partonic form factor
$ \widehat{\widetilde{W}}_{ba} $:
\begin{eqnarray}
 &  & \widetilde{W}_{HA}\left( b,Q,x,z\right) =\sum _{a,b}\, \int _{x}^{1}\frac{d\xi _{a}}{\xi _{a}}\int _{z}^{1}\frac{d\xi _{b}}{\xi _{b}}\nonumber \\
 &  & \times D_{H/b}(\xi _{b},\mu _{F})\widehat{\widetilde{W}}_{ba}\left( b,Q,\xhat ,\zhat ,\mu _{F}\right) \nonumber \\
 &  & \times f_{a/A}(\xi _{a},\mu _{F}),\label{WHA} 
\end{eqnarray}
 where
\begin{eqnarray}
 &  & \wh{\wt{W}}_{ba}(b,Q,\xhat ,\zhat ,\mu _{F})=\sum _{j}e_{j}^{2}\, e^{-{\cal S}(b,Q,C_{1},C_{2})}\nonumber \\
 &  & \times {\mathcal{C}}^{out}_{b/j}\left( \zhat ,\mu _{F}b;\frac{C_{1}}{C_{2}}\right) {\mathcal{C}}_{j/a}^{in}\left( \xhat ,\mu _{F}b;\frac{C_{1}}{C_{2}}\right) .\label{W0} 
\end{eqnarray}
Here $ \xhat \equiv x/\xi _{a},\, \zhat \equiv z/\xi _{b} $. The indices
$ a,b $ in Eq.~(\ref{WHA}) are summed over all quark flavors and gluons;
the summation over $ j $ in Eq.~(\ref{W0}) is over the quarks only.
The fractional charge of a quark $ j $ is denoted as $ e^{2}e_{j}^{2} $.
The parton distributions and fragmentation functions are separated from the
partonic form factor $ \wh{\wt{W}}_{ba} $ at the factorization scale $ \mu _{F} $.
The Sudakov factor $ {\cal S}(b,Q,C_{1},C_{2}) $ is an all-order sum of
logarithms $ \ln ^{m}{(q_{T}^{2}/Q^{2})} $. It is given by an integral
between scales $ C^{2}_{1}/b^{2} $ and $ C^{2}_{2}Q^{2} $ (where $ C_{1} $
and $ C_{2} $ are constants of order 1) of two functions $ {\cal A}\left( \alpha _{S}(\bar{\mu });C_{1}\right)  $
and $ {\cal B}(\alpha _{S}(\bar{\mu });C_{1},C_{2}) $ appearing in the
solution of equations for renormalization- and gauge-group invariance:
\begin{eqnarray}
 &  & {\cal S}=\int _{C_{1}^{2}/b^{2}}^{C_{2}^{2}Q^{2}}\frac{d\bar{\mu }^{2}}{\bar{\mu }^{2}}\Biggl [\ln \left( \frac{C_{2}^{2}Q^{2}}{\bar{\mu }^{2}}\right) {\cal A}\left( \alpha _{S}(\bar{\mu });C_{1}\right) \nonumber \\
 &  & +{\cal B}(\alpha _{S}(\bar{\mu });C_{1},C_{2})\Biggr ].\label{Smassless} 
\end{eqnarray}
The functions $ {\cal C}^{in} $, $ {\cal C}^{out} $ contain perturbative
corrections to contributions from the incoming and outgoing hadronic jets,
respectively. To evaluate the Fourier-Bessel transform integral, $ \widetilde{W}_{HA}\left( b,Q,x,z\right)  $
should be also defined at $ b\gtrsim 1\mbox {\, GeV}^{-1} $, where the
perturbative methods are not trustworthy. The continuation of $ \widetilde{W}_{HA}\left( b,Q,x,z\right)  $
to the large-$ b $ region is realized with the help of some phenomenological
model, as discussed, \emph{e.g.}, 
in Refs.~\cite{Collins:1985kg,Qiu:2000hf,Kulesza:2002rh}.
\begin{figure*}
{\centering \resizebox*{0.9\textwidth}{!}{\includegraphics{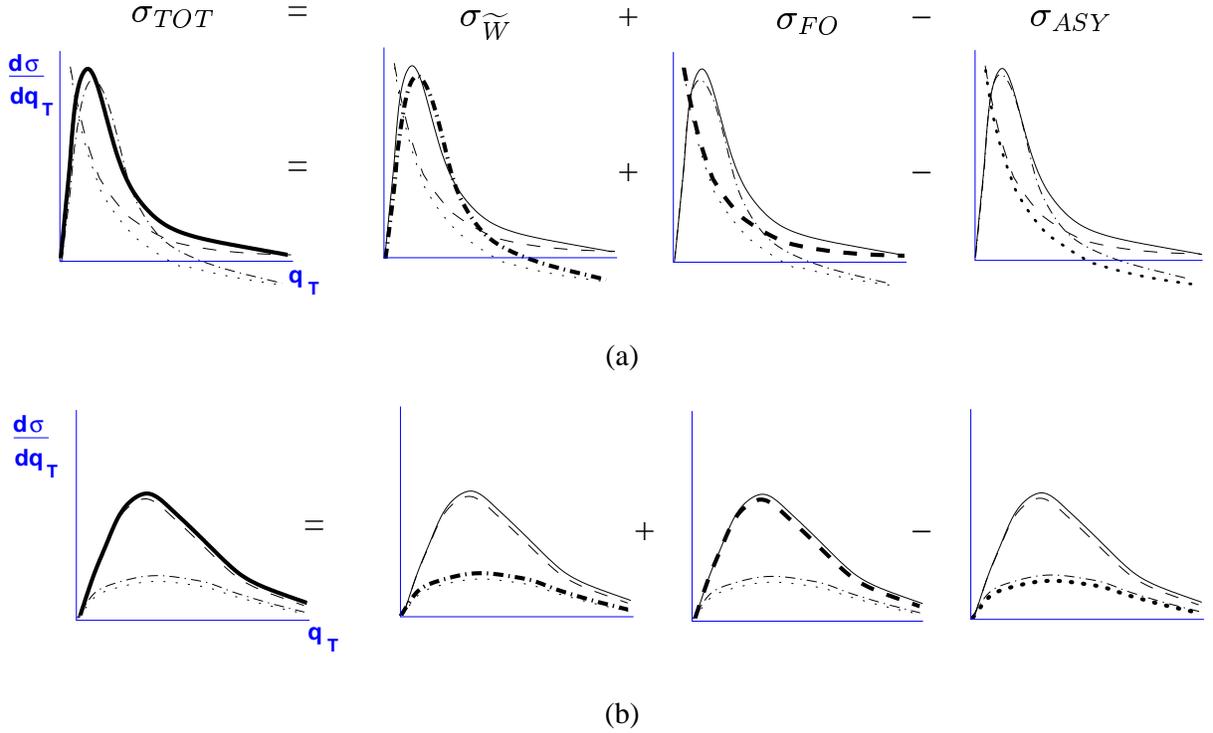}} \par}

\caption{\label{fig:TotWPertAsy}Balance of various terms in the total resummed cross
section \protect$ d\sigma _{TOT}/dq_{T}\protect $: (a) away from the threshold
(\protect$ Q\gg M\protect $); (b) near the threshold (\protect$ Q\approx M\protect $).
In each plot the thick curves correspond to the \char`\"{}active\char`\"{}
cross section (TOT, FO, W or ASY), and the thin curves correspond to the
other three cross sections. }
\end{figure*}

As noted above, the resummed cross section in Eq.~(\ref{Wmassless}), which
we shall label as $ \sw  $, is derived in the limit $ \qt ^{2}\ll Q^{2} $.
In the region $ \qt ^{2}\gtrsim Q^{2} $, the standard finite-order (FO)
perturbative result, $ \sfo  $, is appropriate. While $ \sw  $ and
$ \sfo  $ represent the correct limiting behavior, we cannot simply add
these two terms to obtain the total cross section, $ \stot  $, as we would
be {}``double-counting{}'' the contributions common to both terms.

The solution is to subtract the overlapping contributions between $ \sw  $
and $ \sfo  $. This overlapping contribution (the \emph{asymptotic piece}
$ \sigma _{ASY} $) is obtained by expanding the $ b $-space integral
in $ \sw  $ out to the finite order of $ \sfo  $. Thus, the complete
result is given by 
\begin{equation}
\label{sigmaTOT}
\frac{d\stot }{d\Phi }=\frac{d\sw }{d\Phi }+\frac{d\sfo }{d\Phi }-\frac{d\sasy }{d\Phi }.
\end{equation}
 At small $ q_{T} $, where terms $ \ln ^{m}(q_{T}^{2}/Q^{2}) $ are
large, $ \sfo  $ cancels well with $ \sasy  $, so that the total cross
section is approximated well by the $ b $-space integral: $ \stot \approx \sw  $.
At $ q^{2}_{T}\gtrsim Q^{2} $, where the logarithms are no longer dominant,
the $ b $-space integral $ \sw  $ cancels with $ \sasy  $, so that
the total cross section is equal to $ \sfo  $ up to higher order corrections:
$ \stot \approx \sfo  $. This interplay of $ \sigma _{\widetilde{W}},\, \sigma _{FO}, $
and $ \sigma _{ASY} $ in $ \sigma _{TOT} $ is illustrated in Fig.~\ref{fig:TotWPertAsy}a.

As we will be referring to these different terms frequently throughout the
rest of the paper, let us present a recap of their roles. 

\begin{itemize}
\item $ \sw  $ is the small-$ q_{T} $ resummed term as given by the CSS formalism
in Eq.~(\ref{Wmassless}); sometimes called {}``the CSS term{}'' \cite{Balazs:1997xd}.
This expression contains the all-order sum of large logarithms of the form
$ \ln ^{m}{(q_{T}^{2}/Q^{2})} $, which is presented as a Fourier-Bessel
transform of the $ b $-space form factor $ \widetilde{W}(b,Q,x,z). $
It is a good approximation in the region $ q_{T}^{2}\ll Q^{2} $.
\item $ \sfo  $ is the \emph{finite-order} (FO) term; sometimes called {}``the
perturbative term{}''. It contains the complete perturbative expression
computed to the relevant order of the calculation $ n $. As such, this
term contains logarithms of the form $ \ln ^{m}{(q_{T}^{2}/Q^{2})} $ only
out to $ m=2n-1 $. It also contains terms that are not important in the
limit $ q_{T}^{2}/Q^{2}\rightarrow 0 $, but dominate when $ q^{2}_{T}\sim Q^{2} $.
Hence, it provides a good approximation in the region $ \qt ^{2}\gtrsim Q^{2} $.
\item $ \sasy  $ is the \emph{asymptotic (ASY)} term. It contains the expansion
of $ \sw  $ out to the same order $ n $ as in $ \sigma _{FO} $.
As such, this term contains logarithms of the form $ \ln ^{m}{(q_{T}^{2}/Q^{2})} $
only out to $ m=2n-1 $. It is precisely what is needed to eliminate the
{}``double-counting{}'' between the $ \sw  $ and $ \sfo  $ terms
in Eq.~(\ref{sigmaTOT}).
\item $ \stot  $ is the \emph{total} (TOT) resummed cross section; sometimes
called {}``the resummed term{}''. It is constructed as $ \stot =\sw +\sfo -\sasy  $.
In the region $ q_{T}^{2}\ll Q^{2} $, $ \sasy  $ precisely cancels
the large terms present in the $ \sfo  $ contribution, so that $ \stot \approx \sw  $.
In the region $ q_{T}^{2}\gtrsim Q^{2} $, $ \sasy  $ approximately
cancels the $ \sw  $ term leaving $ \sfo  $ as the dominant representation
of the total cross section: $ \stot \approx \sfo  $. Hence, when calculated
to a sufficiently high order of $ \alpha _{S} $, $ \sigma _{TOT} $
serves as a good approximation at all $ q_{T} $.
\end{itemize}
~In a practical calculation in low orders of PQCD, one may want to further
improve the cancellation between $ \sigma _{\widetilde{W}} $ and $ \sigma _{ASY} $
at $ q_{T}^{2}\gtrsim Q^{2} $. This improvement can be achieved by introducing
a kinematical correction in these terms that accounts for the reduction of
the allowed phase space for the longitudinal variables $ x $ and $ z $
at non-zero $ q_{T} $. The purpose of this kinematical correction is quite
similar to the purpose of the inclusive scaling variable discussed in Subsection~\ref{sec:DISrescalingVariable}:
it removes contributions from the unphysically small $ x $ and $ z $
that make the difference $ \sigma _{\widetilde{W}}-\sigma _{ASY} $ non-negligible
as compared to $ \sigma _{FO} $. Note that the resummed cross sections
with and without the kinematical correction are formally equivalent to one
another up to higher-order corrections. Further discussion of this issue
can be found in Appendix~\ref{Appendix:KinematicalCorrection}, which introduces
the kinematical correction for the resummed heavy-quark $ q_{T} $ distributions.

\section{Extension of the CSS Formalism to heavy-quark production\label{sec:MassiveCSS}}

In the previous Section, we presented a procedure for the resummation of
distributions $ d\sigma /dq_{T}^{2} $ in the limit when $ Q^{2} $ is
much larger than all other momentum scales, $ Q^{2}\gg q_{T}^{2},\, \{m_{q}^{2}\}. $
We now are ready to discuss its extension to the case when the heavy-quark
mass is not negligible. For simplicity, we again assume that only one heavy
flavor $ h $ has the mass $ M $ comparable with $ Q $: $ Q^{2}\sim M^{2}\gg \Lambda _{QCD}^{2} $.
The generalization for several heavy flavors can be realized through the
conventional sequence of factorization subschemes, in which the heavy quarks
become active partons at energy scales above their mass, and are treated
as non-partonic particles at energy scales below their mass. 

We start by rewriting Eq.~(\ref{WHA}) in a form analogous to Eq.~(4.3)
of Ref.~\cite{Collins:1985kg}, where the form factor $ \widetilde{W} $
was given for the Drell-Yan process:
\begin{widetext}
\begin{eqnarray}
 &  & \widetilde{W}_{HA}\left( b,Q,x,z\right) =\sum _{j}e_{j}^{2}\, \overline{\scrP }^{out}_{H/j}\left( z,b,\{m_{q}\},\frac{C_{1}}{C_{2}}\right) \overline{\scrP }_{j/A}^{in}\left( x,b,\{m_{q}\},\frac{C_{1}}{C_{2}}\right) \nonumber \\
 &  & \times \exp \Biggl \{-\int _{C_{1}^{2}/b^{2}}^{C_{2}^{2}Q^{2}}\frac{d\bar{\mu }^{2}}{\bar{\mu }^{2}}\Biggl [\ln \left( \frac{C_{2}^{2}Q^{2}}{\bar{\mu }^{2}}\right) {\cal A}\left( \alpha _{S}(\bar{\mu });\left\{ \frac{\bar{\mu }}{m_{q}}\right\} ;C_{1}\right) +{\cal B}\left( \alpha _{S}(\bar{\mu });\left\{ \frac{\bar{\mu }}{m_{q}}\right\} ;C_{1},C_{2}\right) \Biggr ]\Biggr \}.\label{W43} 
\end{eqnarray}
\end{widetext}

Here the function $ \overline{\scrP }_{j/A}^{in}\left( x,b,\{m_{q}\},C_{1}/C_{2}\right)  $
describes contributions associated with the incoming hadronic jet. As illustrated
in Appendix~\ref{Appendix:Chg}, $ \overline{\scrP }_{j/A}^{in}\left( x,b,\{m_{q}\},C_{1}/C_{2}\right)  $
is related to the $ k_{T} $-dependent parton distribution $ \scrP _{j/A}^{in}\left( x,k_{T},\{m_{q}\}\right)  $.
Similarly, the function $ \overline{\scrP }^{out}_{H/j}\left( z,b,\{m_{q}\},C_{1}/C_{2}\right)  $
describes contributions associated with the outgoing hadronic jet~\cite{Collins:1982va}.
It is related to the $ k_{T} $-dependent fragmentation function $ \scrP ^{out}_{H/j}\left( z,k_{T},\{m_{q}\}\right)  $.
The functions $ {\cal A} $ and $ {\cal B} $ are the same as in Eq.~(\ref{Smassless}),
except that now they retain the dependence on the quark masses $ \{m_{q}\}=m_{u},m_{d},m_{s},...,M $. 

Eq.~(\ref{WHA}) presents a special case of Eq.~(\ref{W43}). It is valid
at short distances, \emph{i.e.}, when $ 1/b $ is much larger than any
of the quark masses $ m_{q} $. In contrast, Eq.~(\ref{W43}) is valid
at all $ b $ \cite{Collins:1985kg}. As shown in Ref.~\cite{Collins:1982uw},
the transition from Eq.~(\ref{W43}) to Eq.~(\ref{WHA}) is possible because
the functions $ \overline{\scrP }_{j/A}^{in} $ and $ \overline{\scrP }^{out}_{H/j} $
factorize when $ b_{0}^{2}/b^{2}\gg \{m_{q}^{2}\}: $ 
\begin{eqnarray}
 &  & \overline{\scrP }_{j/A}^{in}\left( x,b,\{m_{q}\},\frac{C_{1}}{C_{2}}\right) \rightarrow \sum _{a}\int _{x}^{1}\frac{d\xi _{a}}{\xi _{a}}\nonumber \\
 & \times  & {\mathcal{C}}_{j/a}^{in}\left( \xhat ,\mu _{F}b;\frac{C_{1}}{C_{2}}\right) f_{a/A}\left( \xi _{a},\left\{ \frac{\mu _{F}}{m_{q}}\right\} \right) ;\nonumber \\
 &  & \overline{\scrP }^{out}_{H/j}\left( z,b,\{m_{q}\},\frac{C_{1}}{C_{2}}\right) \rightarrow \sum _{b}\int _{z}^{1}\frac{d\xi _{b}}{\xi _{b}}\nonumber \\
 & \times  & D_{H/b}\left( \xi _{b},\left\{ \frac{\mu _{F}}{m_{q}}\right\} \right) {\mathcal{C}}_{b/j}^{out}\left( \wh{z},\mu _{F}b;\frac{C_{1}}{C_{2}}\right) .\label{PbFactorization} 
\end{eqnarray}
Here we introduced a frequently encountered constant $ b_{0}\equiv 2e^{-\gamma _{E}}\approx 1.123 $.
We see that the form-factor $ \wt{W}_{HA} $ is well-defined both for non-zero
quark masses and in the massless limit. Hence, it does not contain negative
powers of the quark masses or logarithms $ \ln \left( m_{q}/Q\right)  $,
with the exception of the collinear logarithms resummed in the parton distributions
and fragmentation functions. 

We will now argue that the factorization rule similar to Eq.~(\ref{PbFactorization})
should also apply in heavy-flavor production when $ M^{2} $ is not negligible
compared to $ b_{0}^{2}/b^{2} $. Indeed, the factorization of the functions
$ \overline{\scrP }_{j/A}^{in} $ and $ \overline{\scrP }^{out}_{H/j} $
in the limit $ b_{0}^{2}/b^{2}\gg \{m_{q}^{2}\} $ \cite{Collins:1982uw}
closely resembles the factorization of the inclusive DIS structure functions
in the limit $ Q^{2}\gg \{m_{q}^{2}\} $ \cite{Amati:1978wx, Amati:1978by, Libby:1978bx, Ellis:1978sf, Ellis:1979ty}.
In both cases the factorization occurs because the dominant contributions
to the cross section come from {}``ladder{}'' cut diagrams with subgraphs
containing lines of drastically different virtualities. More precisely, the
leading regions in such diagrams can be decomposed into hard subgraphs, which
contain highly off-shell parton lines; and quasi-collinear subgraphs, which
contain lines with much lower virtualities and momenta approximately collinear
to $ p_{A}^{\mu } $ (in the case of $ F_{h/A}(x,Q^{2}) $ or $ \overline{\scrP }_{j/A}^{in} $)
or $ p_{H}^{\mu } $ (in the case of $ \overline{\scrP }^{out}_{H/j} $).
In the functions $ \overline{\scrP }_{j/A}^{in} $ and $ \overline{\scrP }^{out}_{H/j} $,
additional soft gluon subgraphs are present, but they eventually do not affect
the proof of the factorization \cite{Collins:1982uw}. The hard subgraphs
contribute to the inclusive coefficient function $ C_{h/a} $ in Eq.~(\ref{F}),
as well as functions $ {\cal C}_{j/a}^{in} $ or $ {\cal C}_{b/j}^{out} $
in Eq.~(\ref{PbFactorization}). The quasi-collinear subgraphs, which are
connected to the hard subgraphs through one on-shell parton on each side
of the momentum cut, contribute to the PDF's (in the inclusive DIS and SIDIS) 
or FF's (in SIDIS). 

The hard subgraphs are characterized by typical transverse momenta $ k_{T}^{2}\gtrsim \mu _{F}^{2}\gg \Lambda _{QCD}^{2}, $
while the PDF's and FF's are characterized by transverse momenta $ k_{T}^{2}\lesssim \mu _{F}^{2} $.
The factorization scale $ \mu _{F} $ is of order $ Q $ in the inclusive
DIS structure functions and $ b_{0}/b $ in the functions $ \overline{\scrP }_{j/A}^{in} $
and $ \overline{\scrP }^{out}_{H/j} $. As discussed in Section~\ref{sec:ACOT},
the factorization in the inclusive DIS can be extended to the case when $ Q $
is comparable to the heavy-flavor mass $ M $, $ Q^{2}\sim M^{2}\gg \Lambda _{QCD}^{2} $.
Given the close analogy between the inclusive DIS structure functions and
the functions $ \overline{\scrP }_{j/A}^{in}, $ $ \overline{\scrP }^{out}_{H/j} $
, it is natural to assume that the latter factorize when $ b_{0}^{2}/b^{2}\sim M^{2}\gg \Lambda _{QCD}^{2} $
as well: 
\begin{eqnarray}
 &  & \overline{\scrP }_{j/A}^{in}\left( x,b,\{m_{q}\},\frac{C_{1}}{C_{2}}\right) =\sum _{a}\int _{x}^{1}\frac{d\xi _{a}}{\xi _{a}}\nonumber \\
 & \times  & {\mathcal{C}}_{j/a}^{in}\left( \xhat ,\mu _{F}b,bM,\frac{C_{1}}{C_{2}}\right) f_{a/A}\left( \xi _{a},\left\{ \frac{\mu _{F}}{m_{q}}\right\} \right) ;\nonumber \\
 &  & \overline{\scrP }^{out}_{H/j}\left( z,b,\{m_{q}\},\frac{C_{1}}{C_{2}}\right) =\sum _{b}\int _{z}^{1}\frac{d\xi _{b}}{\xi _{b}}\nonumber \\
 & \times  & D_{H/b}\left( \xi _{b},\left\{ \frac{\mu _{F}}{m_{q}}\right\} \right) {\mathcal{C}}_{b/j}^{out}\left( \wh{z},\mu _{F}b,bM,\frac{C_{1}}{C_{2}}\right) .\nonumber \\
 &  & \label{PbFactorizationHeavy} 
\end{eqnarray}

The main difference between Eqs.~(\ref{PbFactorization}) and (\ref{PbFactorizationHeavy})
is contained in the functions $ {\cal C}^{in}_{j/a} $ and $ {\mathcal{C}}_{b/j}^{out} $,
which now explicitly depend on $ M. $ These functions can be calculated
according to their definitions given in Ref.~\cite{Collins:1981uk}. The
unrenormalized expressions for the $ {\cal C} $-functions contain ultraviolet
singularities. To cancel these singularities, we introduce counterterms according
to the procedure described in Section~\ref{sec:ACOT}: that is, graphs with
internal heavy-quark lines are renormalized in the $ \overline{MS} $ scheme
if $ \mu _{F}\sim b_{0}/b>M$ and by zero-momentum subtraction if $ b_{0}/b<M $.
This choice leads to the explicit decoupling of diagrams with heavy quark
lines at $ b\gtrsim b_{0}/M $. In particular, the decoupling implies that
contributions to Eq.~(\ref{PbFactorizationHeavy}) with $ j,a, $ or $ b $
equal to $ h $ are power-suppressed at $ b>b_{0}/M. $

We now consider other sources of the dependence on $ M $ in $ d\sigma /d\Phi . $
Firstly, according to Eq.~(\ref{W43}), there is a dependence on $ M $
in the Sudakov functions $ {\cal A}(\alpha _{S}(\bar{\mu });\bar{\mu }/M;C_{1}) $
and $ {\cal B}(\alpha _{S}(\bar{\mu });\bar{\mu }/M;C_{1},C_{2}) $. Due
to the decoupling, the mass-dependent terms in the Sudakov factor vanish
at $ b\gtrsim b_{0}/M $, except for perhaps terms of  truly nonperturbative
nature, like the intrinsic heavy quark component \cite{Brodsky:1980pb}.
As mentioned above, in this paper such nonperturbative component is ignored.
Secondly, there may also be mass-dependent terms \emph{in the finite-order
cross section,} which are not associated with the leading contributions resummed
in the $ \widetilde{W} $-term: those are the terms that contribute to
the remainder in Eq.~(\ref{Wmassless}). The terms of both types are correctly
included in $ d\sigma _{TOT}/d\Phi  $. Indeed, the terms of the first
type appear in all three terms $ d\sigma _{\wt{W}}/d\Phi  $, $ d\sigma _{FO}/d\Phi  $,
and $ d\sigma _{ASY}/d\Phi  $. Two out of three contributions (in $ d\sigma _{\wt{W}}/d\Phi  $
and $ d\sigma _{ASY}/d\Phi  $, or $ d\sigma _{FO}/d\Phi  $ and $ d\sigma _{ASY}/d\Phi  $)
cancel with one another, leaving the third contribution uncancelled in $ d\sigma _{TOT}/d\Phi  $.
The terms of the second type are contained only in $ d\sigma _{FO}/d\Phi  $,
so that they are automatically included in $ d\sigma _{TOT}/d\Phi  $.

The treatment of the massive terms simplifies more if we adapt the S-ACOT
factorization scheme, in which the heavy quark mass is set to zero in the
hard parts of the flavor-excitation subprocesses. As a result, $ M $ is
neglected in the flavor-excitation contributions to the hard cross section
$ \sigma _{FO}, $ asymptotic term $ \sigma _{ASY} $, and $ {\cal C} $-functions
in the $ \widetilde{W} $-term. The mass-dependent terms are further omitted
in the perturbative Sudakov factor $ {\cal S} $. At the same time, all
mass-dependent terms are kept in $ \sigma _{FO} $, $ \sigma _{ASY} $,
and $ {\cal C} $-functions for gluon-initiated subprocesses. 

As we will demonstrate in the next section, in this prescription the cross
section $ \sigma _{TOT} $ resums the soft and collinear logarithms, when
these logarithms are large, and reduces to the finite-order cross section,
when these logarithms are small. In particular, at $ Q\sim M $ the finite-order
flavor-creation terms approximate well the heavy-quark cross section. Hence
we expect that $ \sigma _{TOT} $ reproduces the finite-order flavor-creation
part at $ Q\sim M $ (Fig.~\ref{fig:TotWPertAsy}b). For this to happen,
the flavor-excitation cross section should cancel well with the subtraction
$ \propto \ln (\mu _{F}/M) $ from the flavor-creation cross section; and
$ \sigma _{\wt{W}} $ should cancel well with $ \sigma _{ASY} $.We find
that these cancellations indeed occur in the numerical calculation, so that
at $ Q\approx M $ $ \sigma _{TOT} $ agrees well with the flavor-creation
contribution to $ \sigma _{FO} $. Similarly, $ \sigma _{TOT} $ reproduces
the massless resummed cross section when $ Q\gg M $ (Fig.~\ref{fig:TotWPertAsy}a).
It also smoothly interpolates between the two regions of $ Q $. 

To summarize our method, the total resummed cross section in the presence
of heavy quarks is calculated as 
\begin{equation}
\label{sigmaTot2}
\frac{d\sigma _{TOT}}{d\Phi }=\frac{d\sigma _{\widetilde{W}}}{d\Phi }+\frac{d\sigma _{FO}}{d\Phi }-\frac{d\sigma _{ASY}}{d\Phi },
\end{equation}
 \emph{i.e.}, using the same combination of the $ \widetilde{W} $-term,
finite-order cross section, and asymptotic cross section as in the massless
case. All three terms on the r.h.s. of Eq.~(\ref{sigmaTot2}) are calculated
in the S-ACOT scheme. 
The $ \widetilde{W} $-term is calculated as
\begin{eqnarray}
 &  & \left( \frac{d\sigma (e+A\rightarrow e+H+X)}{d\Phi }\right) _{\wt{W}}=\frac{\sigma _{0}F_{l}}{S_{eA}}\frac{A_{1}(\psi ,\varphi )}{2}\nonumber \\
 &  & \times \int \frac{d^{2}\vec{b}}{(2\pi )^{2}}e^{i\vec{q}_{T}\cdot \vec{b}}\widetilde{W}_{HA}(b,Q,M,x,z),\label{Wmassive} 
\end{eqnarray}
where the form-factor $ \widetilde{W}_{HA}(b,Q,M,x,z) $ is
\begin{widetext}
\begin{eqnarray}
 &  & \widetilde{W}_{HA}\left( b,Q,M,x,z\right) =\sum _{a,b}\, \int _{\chi _{a}}^{1}\frac{d\xi _{a}}{\xi _{a}}\int _{z}^{1}\frac{d\xi _{b}}{\xi _{b}}D_{H/b}\left( \xi _{b},\left\{ \frac{\mu _{F}}{m_{q}}\right\} \right) f_{a/A}\left( \xi _{a},\left\{ \frac{\mu _{F}}{m_{q}}\right\} \right) \nonumber \\
 &  & \times \sum _{j=u,\bar{u},d,\bar{d}...}e_{j}^{2}{\mathcal{C}}^{out}_{b/j}\left( \zhat ,\mu _{F}b,bM;\frac{C_{1}}{C_{2}}\right) {\mathcal{C}}_{j/a}^{in}\left( \frac{\chi _{a}}{\xi _{a}},\mu _{F}b,bM;\frac{C_{1}}{C_{2}}\right) e^{-S_{ba}(b,Q,M)},\label{WHA2} 
\end{eqnarray}
\end{widetext} and
\begin{eqnarray}
 &  & S_{ba}(b,Q,M)\equiv \int _{C_{1}^{2}/b^{2}}^{C_{2}^{2}Q^{2}}\frac{d\ov{\mu }^{2}}{\ov{\mu }^{2}}\nonumber \\
 &  & \times \Biggl [\ASud (\alpha _{S}(\ov{\mu });C_{1})\ln \left( \frac{C_{2}^{2}Q^{2}}{\ov{\mu }^{2}}\right) \nonumber \\
 &  & +\BSud (\alpha _{S}(\ov{\mu });C_{1},C_{2})\Biggr ]+S_{ba}^{NP}(b,Q,M).
\end{eqnarray}
As in the factorization of inclusive DIS structure functions (cf. Section~\ref{sec:ACOT}),
we find it useful to replace Bjorken $ x $ by scaling variables 
\begin{equation}
\label{chih1}
\chi _{h}=x\left( 1+\frac{M^{2}}{z(1-z)Q^{2}}\right) 
\end{equation}
 in $ \sigma _{FO} $ for the flavor-excitation subprocesses, and 
\begin{equation}
\label{chih2}
\chi ^{'}_{h}=x\left( 1+\frac{M^{2}+z^{2}q_{T}^{2}}{z(1-z)Q^{2}}\right) 
\end{equation}
 in $ \sigma _{\widetilde{W}} $ and $ \sigma _{ASY} $. The purpose
of these scaling variables is to enforce the correct threshold behavior of
terms with incoming heavy quarks. Eqs.~(\ref{chih1}) and (\ref{chih2})
are derived in detail in Appendix~\ref{Appendix:KinematicalCorrection}.

\section{\label{sec:PhotonGluon}Massive resummation for photon-gluon fusion}

We now analyze contributions to the total resummed cross section $ d\sigma _{TOT}/d\Phi  $
from the $ {\cal O}(\alpha _{S}^{0}) $ heavy-flavor excitation subprocess
$ \gamma ^{*}(q)+h(p_{a})\rightarrow h(p_{b}) $ (Fig.~\ref{fig:diag2}a)
and $ {\cal O}(\alpha _{S}) $ photon-gluon fusion subprocess $ \gamma ^{*}(q)+G(p_{a})\rightarrow h(p_{b})+\bar{h}(p_{s}) $
(Fig.~\ref{fig:diag2}b). Since we work in the S-ACOT scheme, only the $ {\cal O}(\alpha _{S}) $
fusion subprocess retains the heavy quark mass, so that we concentrate on
that process first. The expression for the $ \gamma ^{*}h $ contribution,
which is the same as in the massless case, is given in Eq.~(\ref{FiniteOrderLO}).
In the following we outline the main results, while details are relegated
to Appendices.

\subsection{Mass-Generalized Kinematical Variables}

Our approach will be to first generalize the kinematical variables from the
massless resummation formalism to {}``recycle{}'' as much of the results
from Refs.~\cite{Meng:1992da,Meng:1996yn,Nadolsky:1999kb} as possible. 

Throughout the derivation, the mass of the incident hadron will be neglected:
$ p_{A}^{2}=0 $. We will use the standard DIS variables $ x,\, Q^{2}, $
and $ z, $ defined by
\begin{eqnarray}
x & \equiv  & \frac{Q^{2}}{2p_{A}\cdot q};\, \, \, Q^{2}\equiv -q^{2};\, \, \, z\equiv \frac{p_{A}\cdot p_{H}}{p_{A}\cdot q}.\label{xHadron} 
\end{eqnarray}
 Since we will be interested in the transverse momentum distributions (or
equivalently, the angular distributions), we next define the transverse momentum
in a frame-invariant manner. The four-vector $ q_{t}^{\mu } $ of the transverse
momentum must be orthogonal to both of the hadrons, so that we have the conditions
$ q_{t}\cdot p_{A}=0 $ and $ q_{t}\cdot p_{H}=0 $. In the massless
case, $ q^{\mu }_{t} $ is simply defined by subtracting off the projections
of the photon's momentum $ q^{\mu } $ onto $ p_{A} $ and $ p_{H} $.
This is slightly modified in the massive case to become
\begin{eqnarray}
q_{t}^{\mu } & = & q^{\mu }-\left( \frac{p_{H}\cdot q}{p_{A}\cdot p_{H}}-M_{H}^{2}\frac{p_{A}\cdot q}{(p_{A}\cdot p_{H})^{2}}\right) p_{A}^{\mu }\nonumber \\
 & - & \frac{p_{A}\cdot q}{p_{A}\cdot p_{H}}p_{H}^{\mu }.
\end{eqnarray}
 Here $ M_{H} $ denotes the mass of the heavy hadron. We find for $ q_{T}^{2}\equiv -q_{t}^{\mu }q_{t\mu }: $
\begin{equation}
\label{qTHadron}
q_{T}^{2}=Q^{2}+2\frac{p_{H}\cdot q}{z}-\frac{M_{H}^{2}}{z^{2}}.
\end{equation}

The kinematical variables at the parton level can be introduced in an analogous
manner. Let $ \xi _{a} $ denote the fraction of the large '$ - $' component
of the incoming hadron's momentum $ p_{A} $ carried by the initial-state parton
$ a $ (\emph{i.e.}, $ \xi _{a}\equiv p_{a}^{-}/p_{A}^{-} $);\footnote{%
We remind the reader that the analysis is performed in the $ \gamma ^{*}A $
c.m.~frame, where the incident hadron moves in the $ -z $ direction.
} and $ \xi _{b} $ denote the fraction of the large '$ + $' component
of the final-state parton's momentum $ p_{b} $ carried by the outgoing hadron
$ H $ (\emph{i.e.}, $ \xi _{b}\equiv p^{+}_{H}/p^{+}_{b} $). We also
assume that $ \xi _{b} $ relates the transverse momenta of $ b $ and
$ H $, as $ (p_{T})_{H}=\xi _{b}(p_{T})_{b}. $ Since all incoming partons
are massless in the S-ACOT factorization scheme, 
we find the following relations
between the hadron-level variables $ x,\, z,\, q_{T} $ and their parton-level
analogs $ \widehat{x},\zhat ,\, \widehat{q}_{T} $:
\begin{eqnarray}
\wh{x} & \equiv  & \frac{Q^{2}}{2\left( p_{a}\cdot q\right) }=\frac{x}{\xi _{a}};\\
\zhat  & \equiv  & \frac{\left( p_{a}\cdot p_{b}\right) }{(p_{a}\cdot q)}=\frac{z}{\xi _{b}};\\
\widehat{q}_{T} & = & q_{T},\label{qthqt} 
\end{eqnarray}
where in the derivation of Eq.~(\ref{qthqt}) we used the first equality
in Eq.~(\ref{qTvscosthetab}).

If we introduce a massive extension of $ \widehat{q}_{T}^{2} $ called
$ \widetilde{q}_{T}^{2} $ and defined by
\begin{equation}
\label{qTtilde}
\wt{q}_{T}^{2}\equiv \widehat{q}^{2}_{T}+\frac{M^{2}}{\zhat ^{2}},
\end{equation}
 then the form of $ \widetilde{q}_{T}^{2} $ in terms of the Lorentz invariants
is identical to the massless case: 
\begin{equation}
\label{wtqT2}
\wt{q}_{T}^{2}=Q^{2}+2\frac{p_{h}\cdot q}{\zhat }.
\end{equation}
We also generalize the usual Mandelstam variables $ \{\shat ,\that ,\uhat \} $
to what we label the {}``mass-dependent{}'' Mandelstam variables $ \{\shat ,\that _{1},\uhat _{1}\} $:
\begin{eqnarray}
\shat  & = & (q+p_{a})^{2},\\
\that _{1} & \equiv  & \that -M^{2}=(q-p_{h})^{2}-M^{2},\\
\uhat _{1} & \equiv  & \uhat -M^{2}=(p_{a}-p_{h})^{2}-M^{2}.
\end{eqnarray}

By using the variables $ \widetilde{q}_{T}^{2} $, $ \shat , $ $ \that _{1} $,
and $ \uhat _{1} $ instead of their counterparts $ q_{T}^{2}, $ $ \shat  $,
$ \that , $ and $ \uhat  $, we shall be able to cast many of the massive
relations in the form of the massless ones. For example, the expressions
for the {}``mass-dependent{}'' Mandelstam variables $ \{\shat ,\that _{1},\uhat _{1}\} $
in terms of the DIS variables can be written as 
\begin{eqnarray}
\shat  & = & Q^{2}\frac{(1-\xhat )}{\xhat };\\
\that _{1} & = & -Q^{2}\frac{\zhat }{\xhat };\\
\uhat _{1} & = & Q^{2}(\zhat -1)-\wt{q}_{T}^{2}\zhat =-Q^{2}\frac{(1-\zhat )}{\xhat }.
\end{eqnarray}
 Note how we made use of the generalized transverse momentum variable $ \widetilde{q}_{T}^{2} $.
These relationships have the same form as their massless counterparts. As
a result, the denominators of the mass-dependent propagators, which are formed
from the invariants $ \shat ,\that _{1} $, and $ \uhat _{1} $, retain
the same form as the denominators of the massless propagators, which are
formed from the invariants $ \shat ,\that , $ and $ \uhat  $.

\subsection{\label{subsec:GammaPframe}Relations between \protect$ \{E_{H},\cos \theta _{H}\}\protect $in
the \protect$ \gamma ^{*}A\protect $ c.m.~frame and \protect$ \{z,\qt ^{2}\}\protect $}

\begin{figure}
{\centering \resizebox*{1\columnwidth}{!}{\includegraphics{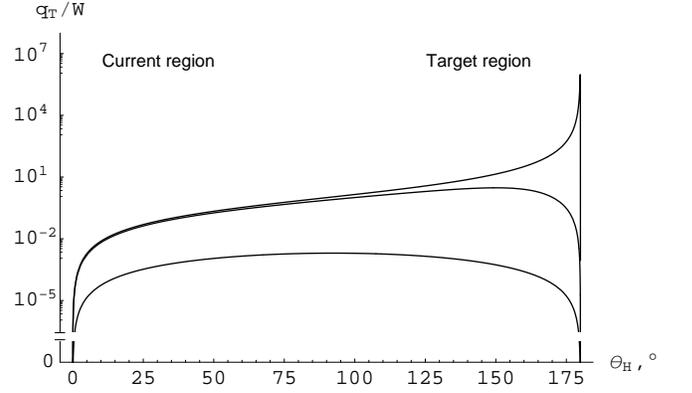}} \par}

\caption{\label{fig:eta_vs_thetaB}Plots of \protect$ q_{T}/W\protect $ vs. \protect$ \theta _{H}\protect $
at various values of \protect$ \lambda \equiv M_{H}/E_{H}=0.999\protect $
(lower curve), \protect$ \lambda =0.5\protect $ (middle curve), and \protect$ \lambda =0.001\protect $
(upper curve).}
\end{figure}

It is useful to convert between the final-state energy $ E_{H} $, polar
angle $ \theta _{H} $ and the Lorentz invariants $ \{z,\qt ^{2}\} $.
Given the $ \gamma ^{*}A $ c.m.~energy $ W^{2}\equiv (q+p_{A})^{2}=Q^{2}(1-\nolinebreak x)/x $
and $ p\equiv |\vec{p}_{H}|=\sqrt{E_{H}^{2}-M_{H}^{2}} $, one easily finds
the following constraints on $ E_{H}, $ $ p $, and $ \cos \theta _{H} $:
\begin{eqnarray}
 & M_{H}\leq E_{H}\leq \frac{W}{2}\left( 1+\frac{M_H^{2}}{W^{2}}\right) , & \label{EHrange} \\
 & 0\leq p\leq \frac{W}{2}\left( 1-\frac{M_H^{2}}{W^{2}}\right) , & \label{prange} 
\end{eqnarray}
and
\begin{equation}
\label{cosThetaHrange}
-1\leq \cos \theta _{H}\leq 1.
\end{equation}

Given $ E_{H} $ and $ \cos \theta _{H} $, we can determine $ z $
and $ q_{T}^{2} $ as
\begin{eqnarray}
z & = & \frac{1}{W}\left( E_{H}+p\cos \theta _{H}\right) ;\label{z} \\
q^{2}_{T} & = & \frac{\left( p^{2}_{T}\right) _{H}}{z^{2}}=W^{2}\frac{p^{2}\left( 1-\cos ^{2}\theta _{H}\right) }{\left( E_{H}+p\cos \theta _{H}\right) ^{2}}.\label{qTvscosthetab} 
\end{eqnarray}
From Eqs.~(\ref{EHrange}-\ref{cosThetaHrange}) the bounds on $ z $
can be found as
\begin{equation}
\frac{M_{H}^{2}}{W^{2}}\leq z\leq 1.
\end{equation}
Note that the first equality in Eq.\textit{\emph{~}}(\textit{\emph{\ref{qTvscosthetab}}})
identifies \textit{\emph{$ q_{T} $}} as the the transverse momentum of
\textit{\emph{$ H $}} rescaled by the final-state fragmentation variable
\textit{\emph{$ z $}}. Hence \textit{\emph{$ q_{T} $}} can be also
interpreted as the leading-order transverse momentum of the fragmenting parton.
Similarly, $\wt q_T = M_T/\zhat$ can be interpreted as the rescaled
transverse mass $M_T$ of the heavy quark.
It also follows from Eqs.~(\ref{z},\ref{qTvscosthetab}) that the two-variable
distribution with respect to the variables $ z $ and $ q_{T} $ coincides
with the two-variable distribution with respect to $ E_{H} $ and 
$ \theta _{H} $:
\begin{equation}
\label{zqt2EcosThetaH}
\frac{d\sigma }{dxdQ^{2}dzdq_{T}}=\frac{d\sigma }{dxdQ^{2}dE_{H}d\theta _{H}}.
\end{equation}
As a result, the distributions in the theoretical variables $ z $ and
$ q_{T} $ are directly related to the distributions in $ E_{H} $ and
$ \theta _{H} $ measured in the experiment. 

Despite the simplicity of the relation (\ref{zqt2EcosThetaH}), $ z $
and $ q_{T} $ are quite complicated functions of $ E_{H} $ and $ \cos \theta _{H} $
individually. This feature is different from the massless case, where there
exists a one-to-one correspondence between $ q_{T} $ and $ \cos \theta _{H} $
for the fixed $ \gamma ^{*}A $ c.m.~energy $ W $: 
\begin{equation}
\label{cosThetaHM0}
\left. \cos \theta _{H}\right| _{M_{H}=0}=\frac{W^{2}-q_{T}^{2}}{W^{2}+q_{T}^{2}}.
\end{equation}
 This relationship does not hold in the massive case, in which \emph{one}
value of $ q_{T} $ corresponds to \emph{two} values of $ \cos \theta _{H} $.
Indeed, Eq.~(\ref{qTvscosthetab}) can be expressed as 
\begin{equation}
\label{qT2W}
\frac{q_{T}^{2}}{W^{2}}=\frac{(1-\lambda ^{2})\left( 1-\cos ^{2}\theta _{H}\right) }{\left( 1+\sqrt{1-\lambda ^{2}}\cos \theta _{H}\right) ^{2}},
\end{equation}
where, according to Eq.~(\ref{EHrange}), the variable $ \lambda \equiv M_{H}/E_{H} $
varies in the following range: 
\begin{equation}
\frac{2M_{H}}{W\left( 1+M_H^{2}/W^{2}\right) }\leq \lambda \leq 1.
\end{equation}
Eq.~(\ref{qT2W}) can be solved for $ \cos \theta _{H} $ as 
\begin{eqnarray}
 &  & \cos \theta _{H}=\frac{1}{\left( q_{T}^{2}+W^{2}\right) \sqrt{1-\lambda ^{2}}}\nonumber \\
 &  & \times \Biggl (-q_{T}^{2}\pm W\sqrt{\left( 1-\lambda ^{2}\right) \left( q_{T}^{2}+W^{2}\right) -q^{2}_{T}}\Biggr ).\label{SolutionscosThetaH} 
\end{eqnarray}
When the energy $ E_{H} $ is much larger than $ M_{H} $ ($ \lambda \rightarrow 0 $)
the solution with the {}``$ + ${}'' sign in Eq.~(\ref{SolutionscosThetaH})
turns into the massless solution (\ref{cosThetaHM0}). The solution with
the {}``$ - ${}'' sign reduces to $ \cos \theta _{H}=-1. $

The physical meaning of the relationship between $ q_{T} $ and $ \cos \theta _{H} $
can be understood by considering plots of $ q_{T}/W $ vs.~$ \theta _{H} $
for various values of $ \lambda  $ (Fig.~\ref{fig:eta_vs_thetaB}). Let
us identify \emph{the current fragmentation region} as that where $ \cos \theta _{H} $
is close to $ +1 $ ($ \theta _{H}=0 $), and \emph{the target fragmentation
region} as that where $ \cos \theta _{H} $ is close to $ -1 $ ($ \theta _{H}=\pi  $).
Firstly, $ q_{T}=0 $ if $ \cos \theta _{H}=1 $ or $ \cos \theta _{H}=-1 $.
Secondly, near the threshold ($ \lambda \rightarrow 1 $) the ratio $ q_{T}/W $
is vanishingly small and symmetric with respect to the replacement of $ \theta _{H} $
by $ (\pi -\theta _{H}) $. Thirdly, as $ \lambda  $ decreases, the
distribution $ q_{T}/W $ vs. $ \theta _{H} $ develops a peak near $ \theta _{H}=180^{\circ } $.
This peak is positioned at $ \cos \theta _{H}=-\sqrt{1-\lambda ^{2}} $,
and its height is $ q_{T}/W=\left( 1-\lambda ^{2}\right) ^{1/2}/\lambda ^{2} $.
For $ \theta _{H}\ll 180^{\circ } $, the distribution rapidly becomes
insensitive to $ \lambda  $; more so for smaller $ \theta _{H} $.

In the limit $ \lambda \rightarrow 0 $, the peak at $ \theta _{H}=180^{\circ } $
turns into a singularity. This singularity resides at the point $ z=0 $
and corresponds to hard diffractive hadroproduction. The analysis of this
region requires diffractive parton distribution functions \cite{Trentadue:1994ka, Berera:1994xh, Graudenz:1994dq, Berera:1996fj, deFlorian:1996fd}
and will not be considered here. For $ \theta _{H}\neq 180^{\circ } $,
one recovers a one-to-one correspondence between $ q_{T}/W $ and $ \cos \theta _{H} $
of the massless case. We see that there is a natural relationship between
$ q_{T} $ and $ \cos \theta _{H} $, which becomes especially simple
in the massless limit. In the following, we concentrate on the limit $ q_{T}\rightarrow 0 $
\emph{and} $ z\neq 0 $, which corresponds to the current fragmentation
region $ \theta _{H}\rightarrow 0 $.

\subsection{Factorized cross sections\label{subsec:FactorizedXSections}}

Next, we consider the factorization of the hadronic cross section. Given
the hadron-level phase space element $ d\Phi \equiv dxdQ^{2}dzdq_{T}^{2}d\varphi  $
and its parton-level analog $ d\wh{\Phi }\equiv d\xhat dQ^{2}d\zhat d\wh{q}_{T}^{2}d\wh{\varphi }, $
all three terms on the r.h.s. of Eq.~(\ref{sigmaTot2}) can be written as
\begin{eqnarray}
 &  & \frac{d\sigma }{d\Phi }=\sum _{a,b}\int _{z}^{1}\frac{d\xi _{b}}{\xi _{b}}\int _{\chi _{a}}^{1}\frac{d\xi _{a}}{\xi _{a}}\nonumber \\
 & \times  & D_{H/b}\left( \xi _{b},\left\{ \frac{\mu _{F}}{m_{q}}\right\} \right) f_{a/A}\left( \xi _{a},\left\{ \frac{\mu _{F}}{m_{q}}\right\} \right) \nonumber \\
 &  & \times \frac{d\wh{\sigma }}{d\wh{\Phi }}\left( \frac{\chi _{a}}{\xi _{a}},\frac{z}{\xi _{b}},\frac{q_{T}}{Q},\frac{\mu _{F}}{Q},\frac{M}{Q}\right) .\label{Factorization} 
\end{eqnarray}

Let us first consider the finite-order cross section $ d\wh{\sigma }_{FO}/d\wh{\Phi } $
. The explicit expression for this cross section at the lepton level can
be found in Appendix~\ref{Appendix:FO}. We are interested in extracting
the leading contribution in this cross section in the limit $ Q\rightarrow \infty  $
with other scales fixed. Specifically, we concentrate on the behavior of
the phase-space $ \delta - $function that multiplies the matrix element
$ \left| {\cal M}\right| ^{2} $:
\begin{eqnarray}
 &  & \frac{d\widehat{\sigma }_{FO}}{d\widehat{\Phi }}\propto \delta \left( \widehat{s}+\widehat{t}+\widehat{u}+Q^{2}-2M^{2}\right) \left| {\mathcal{M}}\right| ^{2}\nonumber \\
 &  & =\delta \left( \widehat{s}+\widehat{t}_{1}+\widehat{u}_{1}+Q^{2}\right) \left| {\mathcal{M}}\right| ^{2}\nonumber \\
 &  & =\delta \left( \left( \frac{1}{\widehat{x}}-1\right) \left( \frac{1}{\widehat{z}}-1\right) -\frac{\wt{q}_{T}^{2}}{Q^{2}}\right) \left| {\mathcal{M}}\right| ^{2}.
\end{eqnarray}
Here we used the mass-generalized variable $ \wt{q}_{T}^{2} $ introduced
in Eq.~(\ref{wtqT2}). Note that in terms of the variables $ \wh{x},\zhat  $,
and $ \widetilde{q}_{T}^{2} $ this expression takes the same form as its
massless version. In the limit $ Q\rightarrow \infty  $, and $ \xhat ,\zhat , $
and $ \wt{q}_{T} $ fixed, the $ \delta  $-function can be transformed
using the relationship
\begin{eqnarray}
\lim _{\varepsilon \rightarrow 0}\delta \left( y_{1}y_{2}-\varepsilon \right)  & \approx  & \frac{\delta (y_{1})}{[y_{2}]_{+}}+\frac{\delta (y_{2})}{[y_{1}]_{+}}\nonumber \\
 & - & \log (\varepsilon )\delta (y_{1})\delta (y_{2}).
\end{eqnarray}
 This transformation yields
\begin{eqnarray}
 &  & \lim _{Q\rightarrow \infty }\delta \left( \widehat{s}+\widehat{t}+\widehat{u}+Q^{2}-2M^{2}\right) \propto \nonumber \\
 &  & \frac{\delta (1-\xhat )}{[1-\zhat ]_{+}}+\frac{\delta (1-\zhat )}{[1-\xhat ]_{+}}\\
 &  & -\log \left( \frac{\widetilde{q}_{T}^{2}}{Q^{2}}\right) \delta \left( 1-\xhat \right) \delta (1-\zhat ).
\end{eqnarray}
This asymptotic expression for the $ \delta  $-function is exactly of
the same form as in the massless case up to the replacement $ \wt{q}_{T}^{2}\rightarrow q_{T}^{2} $.

Furthermore, in the above limit the matrix element $ |{\mathcal{M}}|^{2} $
itself contains singularities when $ Q^{2}\gg \wt{q}_{T}^{2} $. In particular,
the largest structure function $ \wh{V}_{1} $ in the $ \gamma ^{*}G $-fusion
subprocess (cf. Eq.~(\ref{V1jg})) contains contributions proportional to
\begin{equation}
\frac{1}{(M^{2}-\wh{t})(M^{2}-\uhat )}\propto \frac{1}{t_{1}u_{1}}\propto \frac{1}{\widetilde{q}_{T}^{2}},
\end{equation}
and
\begin{equation}
\frac{M^{2}}{\that ^{2}_{1}\uhat ^{2}_{1}}\propto \frac{M^{2}}{\wt{q}_{T}^{4}}=\frac{\zhat ^{4}M^{2}}{\left( \zhat ^{2}q_{T}^{2}+M^{2}\right) ^{2}}.
\end{equation}
When $ M $ is not negligible, these contributions are finite and comparable
with other terms. However, in the limit when \emph{both} $ M $ and $ q_{T} $
are much less than $ Q $, the terms of the first type diverge as $ 1/q_{T}^{2} $.
The terms of the second type vanish at $ q_{T}\neq 0 $ and yield a finite
contribution at $ q_{T}=0 $. These non-vanishing contributions are precisely
the ones that are resummed in the $ \widetilde{W} $-term; in the total
resummed cross section $ \sigma _{TOT} $, they have to be subtracted in
the form of the asymptotic cross section $ \sigma _{ASY} $ to avoid the
double-counting between $ \sigma _{FO} $ and $ \sigma _{\wt{W}} $. 

To precisely identify these terms, we calculate them from their definitions,
as described in Appendix~\ref{Appendix:Chg}. Since the $ {\cal O}(\alpha _{S} $)
$ \gamma ^{*}G $ subprocess is finite in the soft limit, it contributes
only to the function $ {\cal C}^{in}_{h/G}(x,\mu _{F}b,bM) $ and not to
the Sudakov factor. The $ {\cal O}(\alpha _{S}/\pi ) $ coefficient in
this function is
\begin{eqnarray}
 &  & {\mathcal{C}}^{in(1)}_{h/G}(\xhat ,\mu _{F}b,bM)=T_{R}x(1-x)\left( 1+c_{1}(bM)\right) \nonumber \\
 &  & +P^{(1)}_{h/G}(x)\left( c_{0}(bM)-\ln \Bigl (\frac{\mu _{F}b}{b_{0}}\Bigr )\right) 
\end{eqnarray}
if $ \mu _{F}\geq M $, and
\begin{eqnarray}
{\mathcal{C}}^{in(1)}_{h/G}(\xhat ,\mu _{F}b,bM) & = & \left. {\mathcal{C}}^{in(1)}_{h/G}(\xhat ,\mu _{F}b,bM)\right| _{\mu _{F}\geq M}\nonumber \\
 & + & P^{(1)}_{h/G}(\widehat{x})\ln \frac{\mu _{F}}{M}
\end{eqnarray}
if $ \mu _{F}<M $. Here $ P^{(1)}_{h/G}(\xi ) $ is the $ \overline{MS} $
splitting function: $ P_{h/G}^{(1)}(\xi )=T_{R}\left( 1-2\xi +2\xi ^{2}\right) , $
with $ T_{R}=1/2. $ The functions $ c_{0}(bM), $ and $ c_{1}(bM) $
denote the parts of the modified Bessel functions $ K_{0}(bM) $ and $ bM\, K_{1}(bM) $
that vanish when $ b\ll 1/M $. They are defined in Eqs.~(\ref{c0}) and
(\ref{c1}), respectively.

We now have all terms necessary to calculate the combination $ ({\cal C}^{in(0)}_{h/h}\otimes f_{h/A})(x)+({\cal C}^{in(1)}_{h/G}\otimes f_{G/A})(x) $,
which serves as the first approximation to the function $ \overline{\scrP }_{h/A}^{in}\left( x,b,M,C_{1}/C_{2}\right)  $.
We find that this combination possesses two remarkable properties: it smoothly
vanishes at $ \mu ^{2}_{F}=b^{2}_{0}/b^{2}\ll M^{2} $ and is differentiable
with respect to $ \ln (\mu _{F}/M) $ at the point $ \mu _{F}=M $. As
a result, the form factor $ \widetilde{W}(b,Q,x,z) $ for the combined
$ {\cal O}(\alpha _{S}^{0}) $ flavor-excitation and $ {\cal O}(\alpha _{S}^{1}) $
flavor-creation channels is a smooth function at all $ b $, which is strongly
suppressed at $ b^{2}\gg b_{0}^{2}/M^{2} $. The physical consequence is
that, for a sufficiently heavy quark, the $ b $-space integral can be
performed over the large-$ b $ region without introducing an additional
suppression of the integrand by nonperturbative contributions. We use this
feature in Section~\ref{sec:NumericalResults}, where we calculate the resummed
cross section for bottom quark production, which does not depend on the nonperturbative
Sudakov factor.

Finally, by expanding the form-factor $ \widetilde{W}_{HA} $ in a series
of $ \alpha _{S}/\pi  $ and calculating the Fourier-Bessel transform integral
in Eq.~(\ref{Wmassive}), we find the following asymptotic piece for the
$ \gamma ^{*}G $ fusion channel: 
\begin{eqnarray}
 &  & \left( \frac{d\wh{\sigma }(e+G\rightarrow e+h+\bar{h})}{d\wh{\Phi }}\right) _{ASY}=\frac{\sigma _{0}F_{l}}{4\pi S_{eA}}\frac{\alpha _{S}}{\pi }\nonumber \\
 &  & \times A_{1}(\psi ,\varphi )\delta (1-\zhat )\nonumber \\
 &  & \times \left[ \frac{P^{(1)}_{h/G}(\xhat )}{\wh{\wt{q}}_{T}^{2}}+\frac{M^{2}\xhat (1-\xhat )}{\wh{\wt{q}}_{T}^{4}}\right] .\label{sigma_ASY} 
\end{eqnarray}
When $ Q\sim M, $ $ d\wh{\sigma }_{ASY}/d\wh{\Phi } $, which is a regular
function at all $ q_{T} $, cancels well with $ d\wh{\sigma }_{\wt{W}}/d\wh{\Phi }. $
In the limit $ Q\rightarrow \infty  $, $ d\wh{\sigma }_{ASY}/d\wh{\Phi } $
precisely cancels the asymptotic terms that appear in the finite-order cross
section $ d\wh{\sigma }_{FO}/d\wh{\Phi }. $

\section{Numerical Results \label{sec:NumericalResults}}

\begin{figure*}
{\centering \subfigure{\resizebox*{0.49\textwidth}{!}{\includegraphics{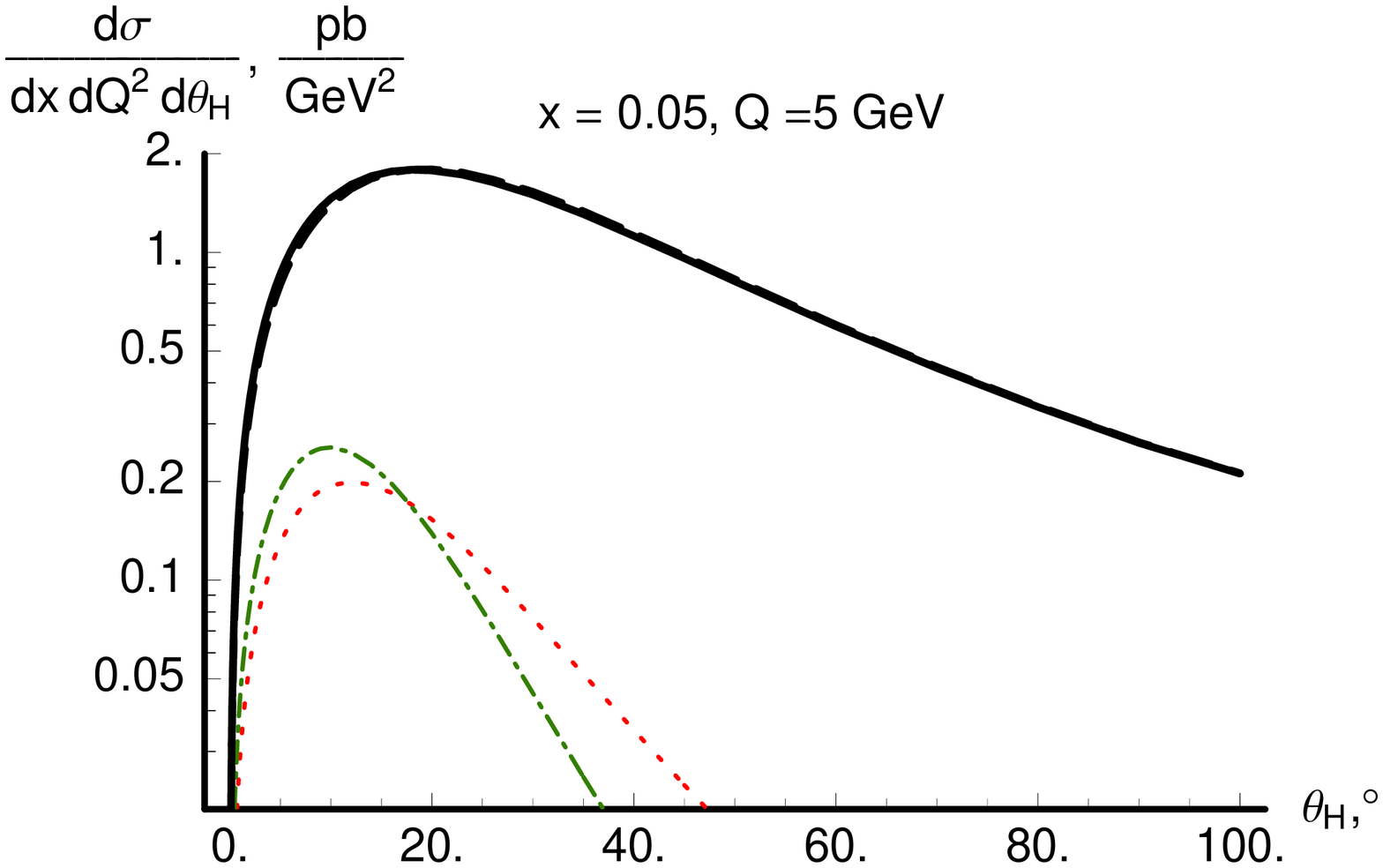}}} \subfigure{\resizebox*{0.49\textwidth}{!}{\includegraphics{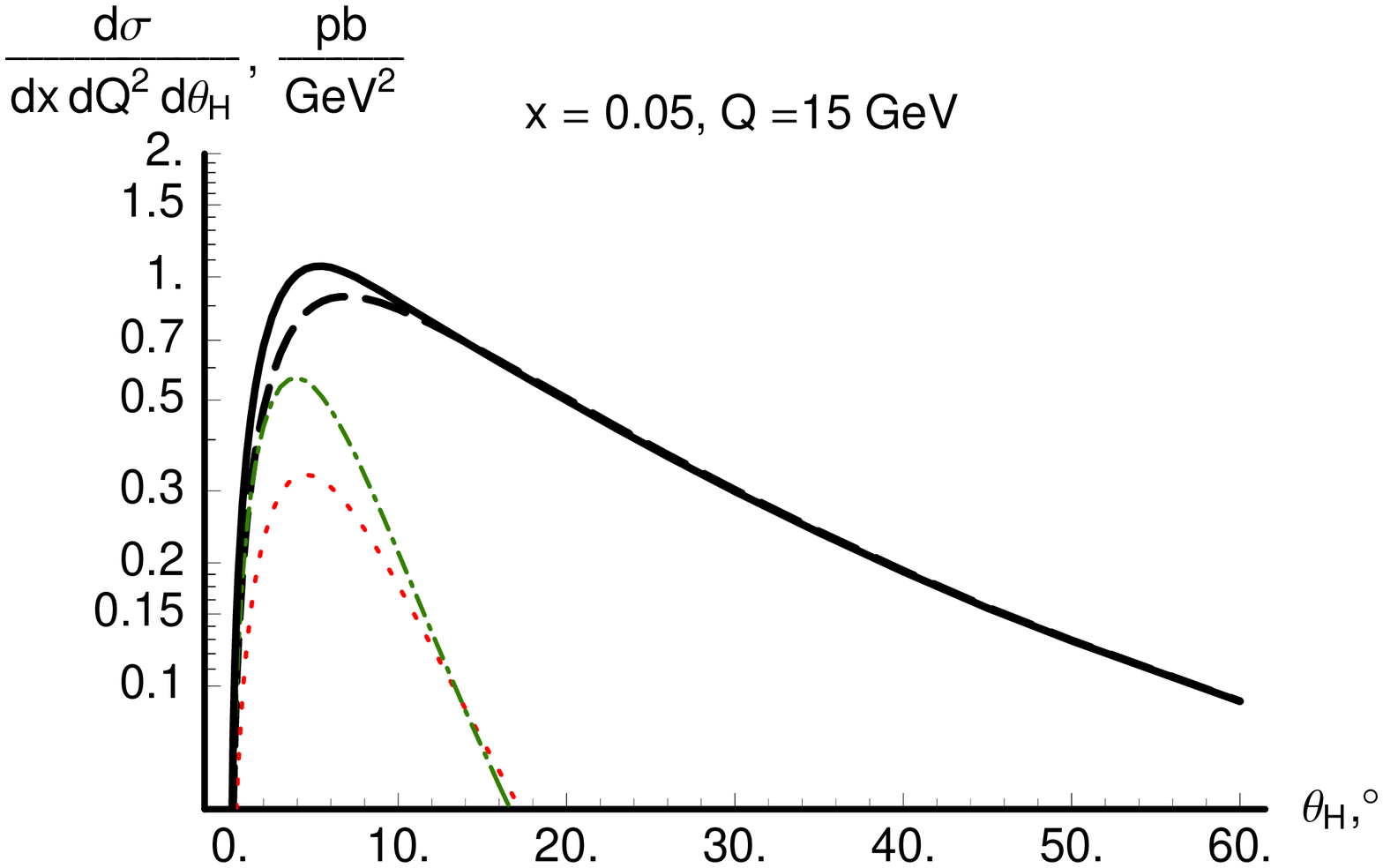}}} \par}

{\centering (a)\hspace{3in}(b)\par}

{\centering \resizebox*{0.495\textwidth}{!}{\includegraphics{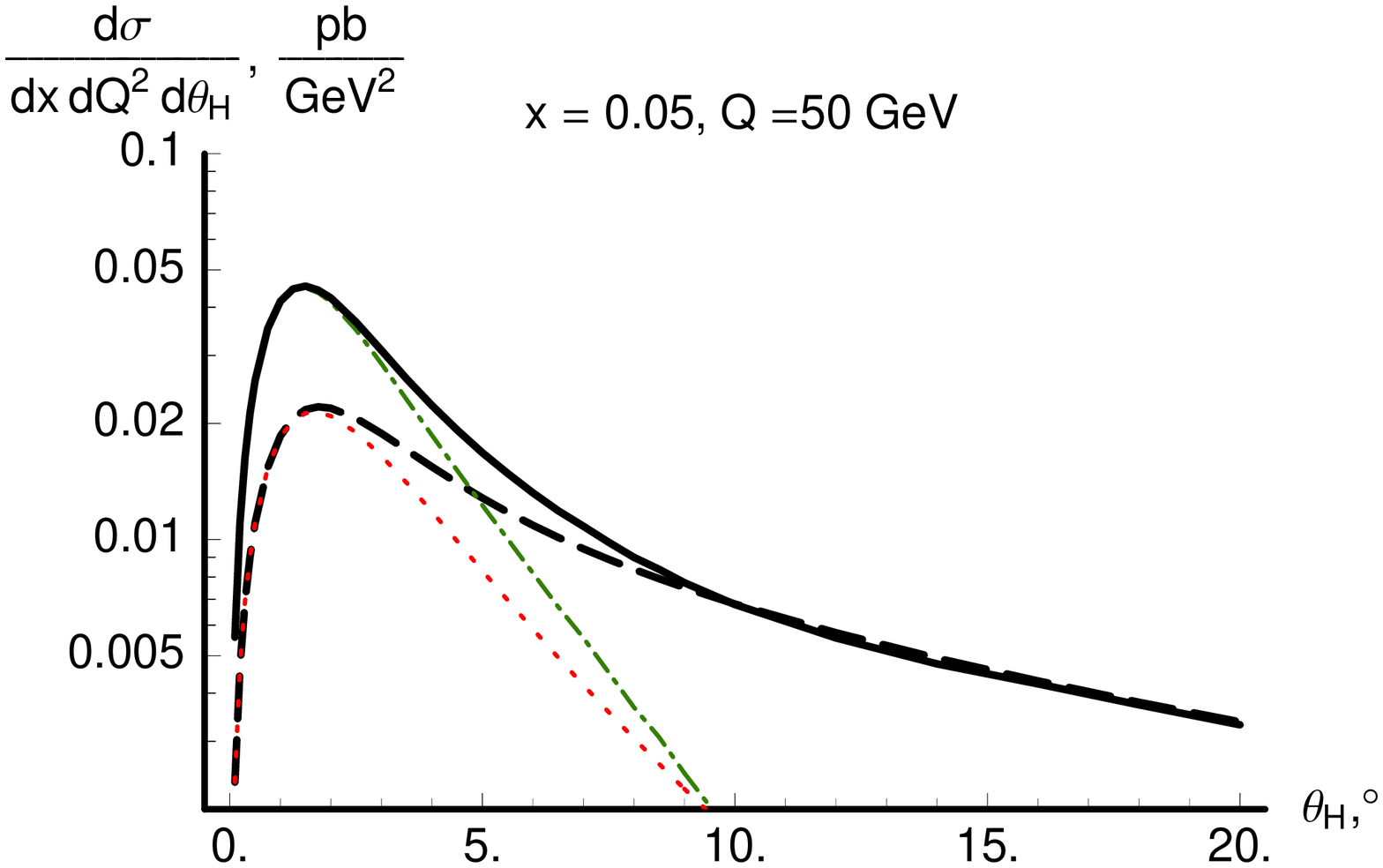}} \resizebox*{0.495\textwidth}{!}{\includegraphics{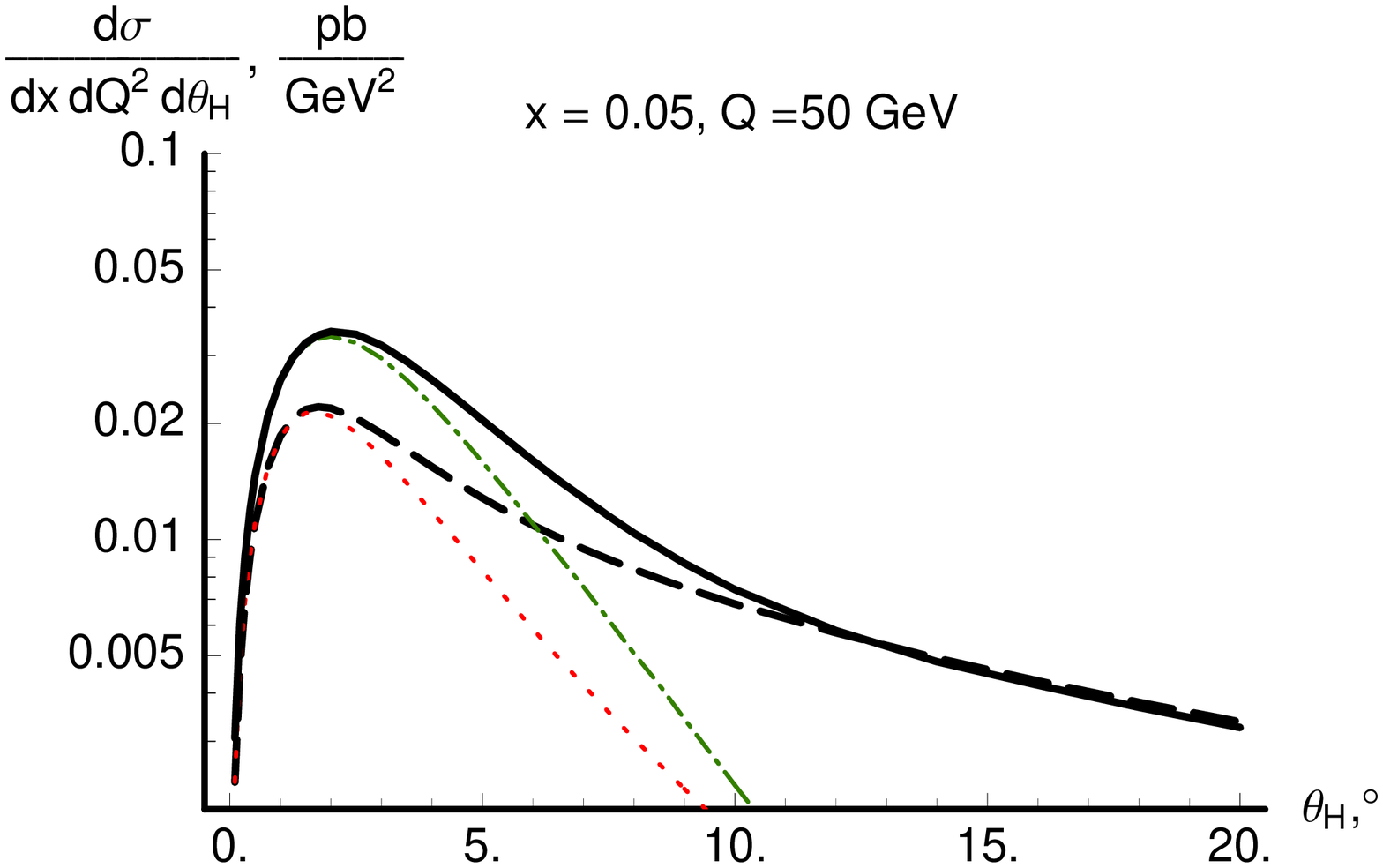}} \par}

{\centering (c)\hspace{3in}(d)\par}

\caption{\label{fig:NumericalResults}The angular distributions of the
bottom hadrons
in the \protect$ \gamma ^{*}p\protect $ c.m.~frame at (a) \protect$ Q=5\protect $
GeV, (b) \protect$ Q=15\protect $ GeV, (c) \protect$ Q=50\protect $
GeV without the Sudakov factor, and (d) \protect$ Q=50\protect $ GeV with
the Sudakov factor. At \protect$ Q=50\protect $ GeV, an additional cut
\protect$ E_{H}>0.1(W/2)\protect $ is made to suppress contributions at
\protect$ z<0.1\protect $, \emph{i.e.}, from the region where the conventional
factorization may be inapplicable. The plots show the finite-order cross
section \protect$ \sigma _{FO}\protect $ (long-dashed line), the \protect$ b\protect $-space
integral \protect$ \sigma _{\wt{W}}\protect $ (dot-dashed line), the asymptotic
piece \protect$ \sigma _{ASY}\protect $ (dotted line), and the full resummed
cross section \protect$ \sigma _{TOT}\protect $ (solid line). }
\end{figure*}
In this Section, we apply the resummation formalism to the production of
bottom quarks at HERA. The calculation is done for the electron-proton c.m.~energy
of $ 300 $ GeV and bottom quark mass $ M=4.5 $ GeV. For simplicity
we assume that the masses of the $ B $-hadrons coincide with the mass
of the bottom quark $ M $. We also neglect the mixing of photons with
$ Z^{0} $-bosons at large $ Q $.

In the following, we discuss polar angle distributions in the $ \gamma ^{*}p $
frame for $ x=0.05 $ and various values of $ Q $. The cross section
is calculated in the lowest-order approximation as discussed in Section~\ref{sec:PhotonGluon}.\footnote{%
The generalization of our approach to higher orders is straightforward. The
next-order calculation should include the $ {\cal O}(\alpha _{S}) $ flavor-excitation
and $ {\cal O}(\alpha ^{2}_{S}) $ flavor-creation channels, which should
appear together to ensure the smoothness of the form factor $ \wt{W}(b,Q,M,x,z) $
and its suppression at $ b\gtrsim 1/M $. 
} The calculation was realized using the CTEQ5HQ PDF's \cite{Lai:1999wy}
and Peterson fragmentation functions \cite{Peterson:1983ak} with $ \varepsilon =0.0033 $
\cite{BottomDISH1}. The finite-order cross section $ d\sigma _{FO}/d\Phi  $
and asymptotic cross section $ d\sigma _{ASY}/d\Phi  $ were calculated
at the factorization scale $ \mu _{F}=Q $. The scale-related constants
in the $ \widetilde{W} $-term were chosen to be $ C_{1}=2e^{-\gamma _{E}}=b_{0} $
and $ C_{2}=1 $, and the factorization scale was $ \mu _{F}=b_{0}/b $.
The $ \widetilde{W} $-term included the $ {\cal O}(\alpha _{S}^{0}) $
$ {\cal C} $-functions $ {\mathcal{C}}^{in(0)}_{h/h}(\xhat ,\mu _{F}b,C_{1}/C_{2}) $,
$ {\mathcal{C}}^{out(0)}_{h/h}(\zhat ,\mu _{F}b,C_{1}/C_{2}) $ and $ {\cal O}(\alpha _{S}^{1}) $
initial-state function $ {\mathcal{C}}^{in(1)}_{h/G}(\xhat ,\mu _{F}b,bM) $.
In addition, it included the perturbative Sudakov factor (\ref{Smassless}),
unless stated otherwise. The Sudakov factor was evaluated at order $ {\cal O}(\alpha _{S}) $,
which was sufficient for this calculation given the order of other terms.
The functions in the Sudakov factor were evaluated as
\begin{equation}
{\cal A}(\mu ;C_{1})=C_{F}\frac{\alpha _{S}(\mu )}{\pi },
\end{equation}
and
\begin{equation}
\BSud (\mu ;C_{1},C_{2})=-\frac{3C_{F}}{2}\frac{\alpha _{S}(\mu )}{\pi }.
\end{equation}

According to the discussion in Section~\ref{sec:PhotonGluon}, our calculation
ignores unknown nonperturbative contributions in the $ \wt{W} $-term.
In the numerical calculation, we also need to define the behavior of the
light-quark PDF's at scales $ \mu _{F}=b_{0}/b<1\mbox {\, GeV} $. Due
to the strong suppression of the large-$ b $ region by the $ M $-dependent
terms in the $ {\cal C} $-functions (cf.~the discussion after Eq.~(\ref{PinhG1})),
the exact procedure for the continuation of the PDF's to small $ \mu _{F} $
has a small numerical effect. We found it convenient to {}``freeze{}''
the scale $ \mu _{F} $ at a value of about $ 1 $ GeV by introducing
the variable $ b_{*}=b/\sqrt{1+\left( b/b_{max}\right) ^{2}} $ \cite{Collins:1985kg}
with $ b_{max}=b_{0}\mbox {\, GeV}^{-1}\approx 1.123\mbox {\, GeV}^{-1} $.
Other procedures \cite{Qiu:2000hf,Kulesza:2002rh} for continuation of $ \wt{W}_{HA}(b,Q,x,z) $
to large values of $ b $ may be used as well. Due to the small sensitivity
of the resummed cross section to the region $ b^{2}\gg b_{0}^{2}/M^{2} $,
all these continuation procedures should yield essentially identical predictions.

\begin{figure}
{\centering \resizebox*{1\columnwidth}{!}{\includegraphics{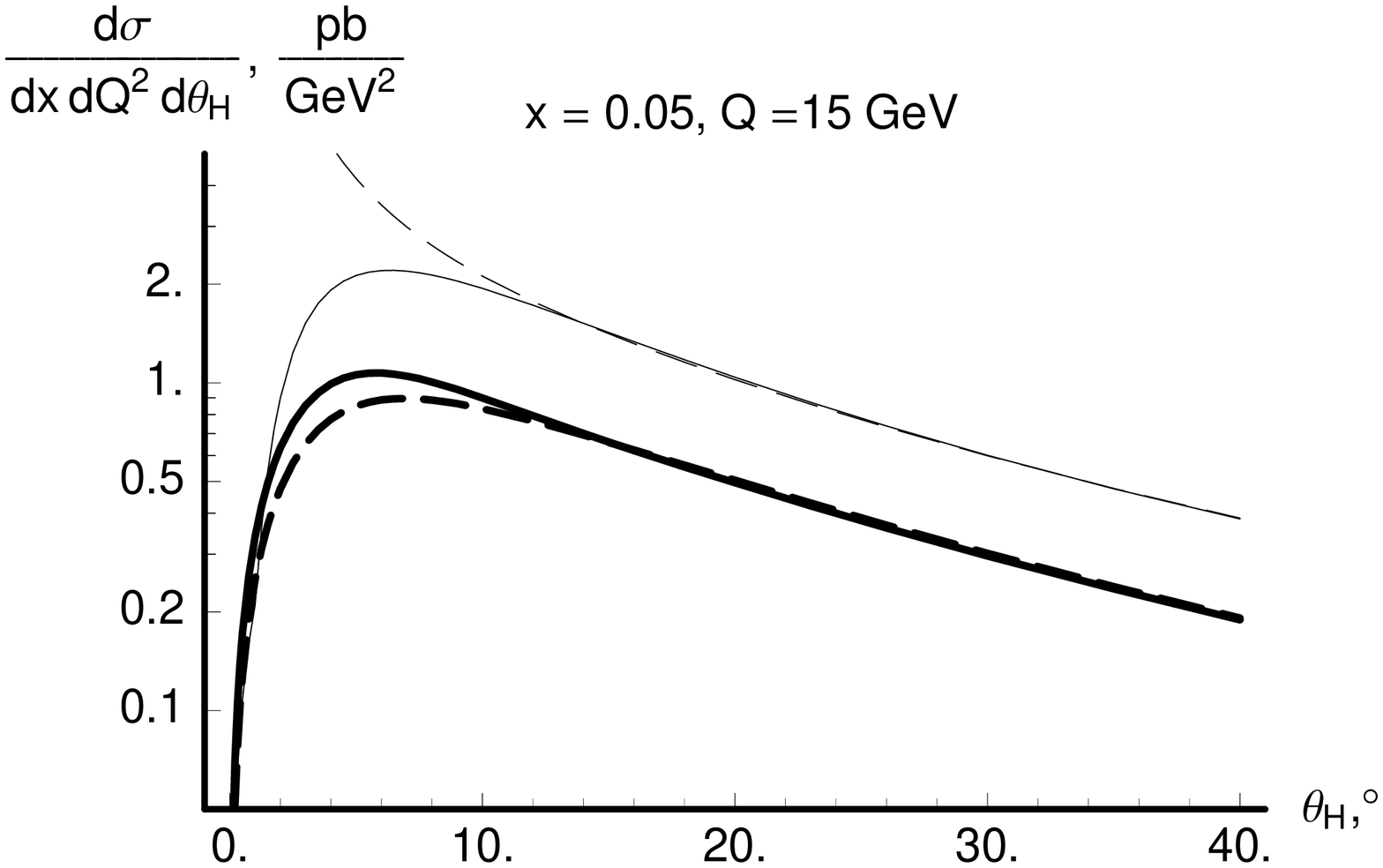}} 
(a)
\resizebox*{1\columnwidth}{!}{\includegraphics{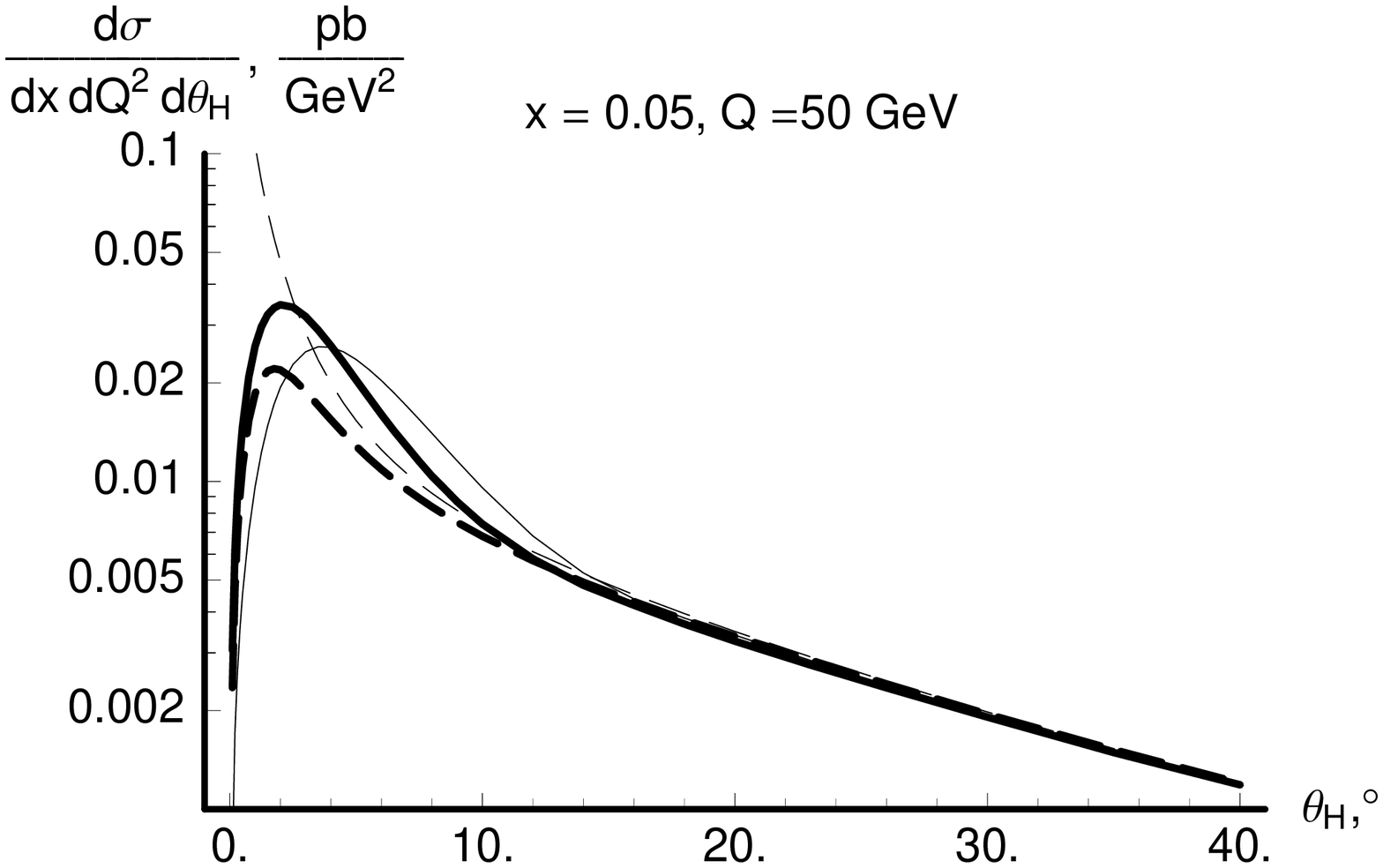}} 
(b)\par}

\caption{\label{fig:MassiveVsMassless} Comparison of the massive and massless cross
sections at (a) \protect$ Q=15\protect $ GeV, and (b) \protect$ Q=50\protect $
GeV. The plots show the massive resummed cross section \protect$ \sigma _{TOT}\protect $
(thick solid line); the massless resummed cross section \protect$ \sigma _{TOT}\protect $
(thin solid line); the massive finite-order cross section \protect$ \sigma _{FO}\protect $
(thick dashed line); and the massless finite-order cross section \protect$ \sigma _{FO}\protect $
(thin dot-dashed line). }
\end{figure}

Fig.~\ref{fig:NumericalResults} demonstrates how various terms in Eq.~(\ref{sigmaTot2})
are balanced in an actual numerical calculation. Near the threshold ($ Q=5 $
GeV, Fig.~\ref{fig:NumericalResults}a) the cross section $ d\sigma _{TOT}/(dxdQ^{2}d\theta _{H}) $
should be well approximated by the $ {\mathcal{O}}(\alpha _{S}) $ flavor-creation
diagram $ \gamma ^{*}+G\rightarrow h+\bar{h} $. We find that this is indeed
the case, since the $ \wt{W} $-term, which does not contain large logarithms,
cancels well with its perturbative expansion $ d\sigma _{ASY}/d\Phi  $.
As a result, the full cross section is practically indistinguishable from
the finite-order term.

At higher values of $ Q, $ we start seeing deviations from the finite-order
result. Fig.~\ref{fig:NumericalResults}b shows the differential distribution
at $ Q=15 $ GeV, \emph{i.e.}, approximately at $ Q^{2}/M^{2}=10 $.
At this energy, $ d\sigma _{TOT}/(dxdQ^{2}d\theta _{H}) $ still agrees
with $ d\sigma _{FO}/(dxdQ^{2}d\theta _{H}) $ at $ \theta _{H}\gtrsim 10^{\circ }, $
but is above $ d\sigma _{FO}/(dxdQ^{2}d\theta _{H}) $ at $ \theta _{H}\lesssim 10^{\circ }. $
The excess is due to the difference $ d\sigma _{\wt{W}}/(dxdQ^{2}d\theta _{H})-d\sigma _{ASY}/(dxdQ^{2}d\theta _{H}), $
\emph{i.e.}, due to the higher-order logarithms. 

Away from the threshold ($ Q=50 $ GeV), $ d\sigma _{TOT}/(dxdQ^{2}d\theta _{H}) $
is substantially larger than the finite-order term at $ \theta _{H}\lesssim 10^{\circ } $,
where it is dominated by $ d\sigma _{\wt{W}}/(dxdQ^{2}d\theta _{H}) $.
In this region, $ d\sigma _{FO}/(dxdQ^{2}d\theta _{H}) $ is canceled well
by $ d\sigma _{ASY}/(dxdQ^{2}d\theta _{H}) $. Note, however, that contrary
to the experience from the massless case, $ d\sigma _{FO}/(dxdQ^{2}d\theta _{H}) $
and $ d\sigma _{ASY}/(dxdQ^{2}d\theta _{H}) $ are not singular at $ \theta _{H}\rightarrow 0 $
due to the regularizing effect of the heavy quark mass in the heavy-quark
propagator at $ \theta _{H}\lesssim 3^{\circ } $. Figs.~\ref{fig:NumericalResults}c
and \ref{fig:NumericalResults}d also compare the distributions with and
without the $ {\cal O}(\alpha _{S}^{1}) $ perturbative Sudakov factor,
respectively. Note that at the threshold the flavor-excitation terms responsible
for $ {\cal S} $ are of a higher order as compared to the $ {\cal O}(\alpha _{S}^{0}) $
flavor-excitation and $ {\cal O}(\alpha _{S}^{1}) $ flavor-creation terms.
Correspondingly, near the threshold the impact of $ {\cal S} $ is expected
to be minimal. This expectation is supported by the numerical calculation,
in which the difference between the curves with and without the $ {\cal O}(\alpha _{S}) $
perturbative Sudakov factor is negligible at $ Q=5 $ GeV, and is less
than a few percent and $ Q=15 $ GeV. In contrast, at $ Q=50 $ GeV the
distribution with the $ {\cal O}(\alpha _{S}) $ Sudakov factor is noticeably
lower and broader than the distribution without it: at some values of $ \theta _{H} $,
the difference in cross sections reaches $ 40\% $. The influence of the
Sudakov factor on the integrated rate is mild: the inclusive cross section
$ d\sigma /(dxdQ^{2}) $ calculated without and with the $ {\cal O}(\alpha _{S}) $
Sudakov factor is equal to $ 330 $ and $ 320\mbox {\, fb/GeV}^{2} $,
respectively. Due to the enhancement at small $ \theta _{H} $, these resummed
inclusive cross sections are larger than the finite-order rate 
$ d\sigma _{FO}/(dxdQ^{2})\approx 260\mbox {\, fb/GeV}^{2} $
by about $ 25\% $. 

It is interesting to compare our calculation with the massless approximation
for the $ \gamma ^{*}G $ contribution. Fig.~\ref{fig:MassiveVsMassless}
shows the finite-order and resummed cross sections calculated in the massive
and massless approaches. In contrast to the massive $ \sigma _{TOT} $,
the massless $ \sigma _{TOT} $ must include the nonperturbative Sudakov
factor $ S^{NP} $, which is not known \emph{a priori} and is usually found
by fitting to the data. To have some reference point, we plot the massless
$ \sigma _{TOT} $ with $ S^{NP}(b)=b^{2}M^{2}/b_{0}^{2}\approx 16b^{2} $,
so that, in analogy to the massive case, the region of $ b\gtrsim b_{0}/M $
in the massless $ \wt{W}(b,Q,x,z) $ is suppressed. Since the heavy-quark
mass has other effects on the shape of $ \wt{W}(b,Q,x,z) $ besides the
cutoff in the $ b $-space, we expect the shape of the massless and massive
resummed curves be somewhat different. This feature is indeed supported by
Fig.~\ref{fig:MassiveVsMassless}b, where at small $ \theta _{H} $ both
resummed curves are of the same order of magnitude, but differ in detail.
Furthermore, the shape of the massless $ \sigma _{TOT} $ can be varied
by adjusting $ S^{NP} $. At the same time, the massive resummed cross
section is uniquely determined by our calculation. 

At sufficiently large $ \theta _{H} $, both the massless and massive resummed
cross sections reduce to their corresponding finite-order counterparts. The
massless cross section significantly overestimates the massive result near
the threshold and at intermediate values of $ Q $. For instance, at $ Q=15 $
GeV (Fig.~\ref{fig:MassiveVsMassless}a) the massless cross section is 
several times larger than the massive cross section in the whole range of $ \theta _{H}. $
In contrast, at $ Q=50 $ GeV (Fig.~\ref{fig:MassiveVsMassless}b) the
massless $ \sigma _{FO} $ agrees well with the massive $ \sigma _{FO} $
at $ \theta _{H}\gtrsim 20^{\circ } $ and overestimates the massive $ \sigma _{FO} $
at $ \theta _{H}\lesssim 20^{\circ }. $ The massive $ \sigma _{TOT} $
is above the massless $ \sigma _{FO} $ at $ 3^{\circ }\lesssim \theta _{H}\lesssim 10^{\circ } $
and below it at $ \theta _{H}\lesssim 3^{\circ }. $ 

The presence of two critical angles ($ \theta _{H}\sim 3^{\circ } $ and
$ \sim 10^{\circ } $) in $ \sigma _{TOT} $ can be qualitatively understood
from the following argument. The rapid rise of $ \sigma _{TOT} $ over
the massive $ \sigma _{FO} $ begins when the small-$ q_{T} $ logarithms
$ \ln ^{m}\left( \wt{q}_{T}^{2}/Q^{2}\right)  $ become large --- say,
when $ \wt{q}_{T}^{2} $ is less than one tenth of $ Q^{2} $. Given
that the Peterson fragmentation function peaks at about $ z\sim 0.95 $,
and that $ Q=50 $ GeV, $ M=4.5 $ GeV, the condition $ \wt{q}_{T}^{2}\sim \, 0.1\, Q^{2} $
corresponds to $ \theta _{H}\sim 8^{\circ }, $ which is close to the observed
critical angle of $ 10^{\circ }. $ Note that in that region $ q_{T}^{2}\gg M^{2}/z^{2}. $
On the other hand, when $ q_{T}^{2} $ is of order $ M^{2}/z^{2} $,
the growth of the logarithms $ \ln \left( \wt{q}_{T}^{2}/Q^{2}\right)  $
is inhibited by the non-zero mass term $ M^{2}/z^{2} $ in $ \wt{q}_{T}^{2} $.
The condition $ q_{T}^{2}\sim M^{2}/z^{2} $ corresponds to $ \theta _{H}\sim 2.5^{\circ } $,
which is approximately where the mass-dependent cross section turns down.

\section{Conclusion and outlook}

In this paper, we presented a method to describe polar angle distributions
in heavy quark production in deep inelastic scattering. This method is realized
in the simplified ACOT factorization scheme \cite{Collins:1998rz,Kramer:2000hn}
and uses the impact parameter space ($ b $-space) formalism \cite{Collins:1981uk,Collins:1982va,Collins:1985kg}
to resum transverse momentum logarithms in the current fragmentation region.
We discussed general features of this formalism and performed an explicit
calculation of the resummed cross section for the $ {\cal O}(\alpha _{S}^{0}) $
flavor-excitation and $ {\cal O}(\alpha _{S}^{1}) $ flavor-creation subprocesses
in bottom quark production. According to the numerical results in Section~\ref{sec:NumericalResults},
the multiple parton radiation effects in this process become important at
$ Q\gtrsim 15 $ GeV (or approximately at $ Q^{2}/M^{2}\gtrsim 10 $).
At $ Q=50 $ GeV, the multiple parton radiation increases the inclusive
cross section by about $ 25\% $ as compared to the finite-order flavor-creation
cross section.

Many aspects of the resummation in the presence of the heavy quarks are similar
to those in the massless resummation. In particular, it is possible to organize
the calculation in the massive case in a close analogy to the massless case
by properly redefining the Lorentz invariants (in particular, by replacing
the Lorentz-invariant transverse momentum $ q_{T} $ in the logarithms
by the rescaled transverse mass 
$ \wt{q}_{T}=\sqrt{q_{T}^{2}+M^{2}/\zhat ^{2}} $). The total resummed
cross section is presented as a sum of the $ b $-space integral $ \sigma _{\wt{W}} $
and the finite-order cross section $ \sigma _{FO} $, from which we subtract
the asymptotic piece $ \sigma _{ASY} $. Constructed in this way, the resummed
cross section reduces to the finite-order cross section at $ Q\approx M $
and reproduces the massless resummed cross section at $ Q\gg M $. 

At the same time, there are important differences between the light- and
heavy-hadron cases. For instance, the light hadron production is sensitive
to the coherent QCD radiation with a wavelength of order $ 1/\Lambda _{QCD} $,
which is poorly known and has to be modeled by the phenomenological {}``nonperturbative
Sudakov factor{}''. In contrast, in the heavy-hadron case such long-distance
radiation is suppressed by the large value of $ M $. Hence, for a sufficiently
heavy $ M $, as in bottom quark production, the resummed cross section
can be calculated without introducing the nonperturbative large-$ b $
contributions. It will be interesting to test the hypothesis about the absence
of such long-distance contributions experimentally. Given the size of the
differential cross sections obtained in Section~\ref{sec:NumericalResults},
accurate tests of this approach will be feasible once the integrated luminosity
of the HERA II run approaches $ 1\mbox {\, fb}^{-1} $. The same calculation
can be done for charm production. However, in that case the region $ b\gtrsim 1\, \mbox {GeV}^{-1} $
is not as suppressed, and the nonperturbative Sudakov factor has to be included. 

\textcolor{black}{Another important improvement in our calculation is more
accurate treatment of threshold effects in fully differential cross sections.
The accuracy in the threshold region is improved by introducing scaling variables
(\ref{chih1}) and (\ref{chih2}) in finite-order and resummed differential
cross sections. These scaling variables generalize the scaling variable proposed
in Ref.~\cite{Tung:2001mv} for inclusive structure functions. They lead
to stable theoretical predictions at all values of $ Q $ and agreement
with the massless result at high energies. }

The extension of our calculation to higher orders is feasible in the near
future, since many of its ingredients are already available in the literature
\cite{Harris:1995tu,Nadolsky:1999kb,Amundson:2000vg}. Furthermore, in a
forthcoming paper we will study the additional effects of threshold resummation
\cite{Kidonakis:1996aq, Kidonakis:1997gm, Kidonakis:1998bk, Kidonakis:1998nf, Kidonakis:1999ze, Kidonakis:2000ui}
in DIS heavy-quark production, so that both transverse momentum and threshold
logarithms are taken into account. We conclude that the combined resummation
of the mass-dependent logarithms $ \ln (M^{2}/Q^{2}) $ and transverse
momentum logarithms $ \ln (q_{T}^{2}/Q^{2}) $ is an important ingredient
of the theoretical framework that aims at matching the growing precision
of the world heavy-flavor data.

\section*{Acknowledgements}

Authors have benefited from discussions with J.~C.~Collins, J.~Smith,
D.~Soper, G.~Sterman, W.-K.~Tung, and other members of the CTEQ Collaboration.
We also appreciate discussions of related topics with A.~Belyaev, B.~Harris,
R.~Vega, W.~Vogelsang, and S.~Willenbrock. We thank S.~Kretzer for the
correspondence on the scaling variable in the ACOT factorization scheme and
R.~Scalise for the participation in early stages of the project. The work
of P.~M.~N. and F.~I.~O. was supported by the U.S. Department of Energy,
National Science Foundation, and Lightner-Sams Foundation. The research of
N. K. has been supported by a Marie Curie Fellowship of the European Community
programme ``Improving Human Research Potential'' under contract number HPMF-CT-2001-01221.
The research of C.-P. Y. has been supported by the National Science Foundation
under grant PHY-0100677.

\appendix

\section{\label{Appendix:Chg}\label{Appendix:Wterm}Calculation of the mass-dependent
\protect$ {\cal C}\protect $-function}

In this Appendix, we derive the $ {\cal O}(\alpha _{S}) $ part of the
function $ {\cal C}^{in}_{h/G}(x,\mu _{F}b,bM) $. This is the only $ {\cal O}(\alpha _{S}) $
term in the heavy-quark $ \widetilde{W} $-term that explicitly depends
on the heavy-quark mass~$ M $. This function appears in the factorized
small-$ b $ expression for the {}``$ b $-dependent PDF{}'' $ \overline{\scrP }_{h/A}^{in}\left( x,b,\{m_{q}\},C_{1}/C_{2}\right)  $:
\begin{eqnarray}
 &  & \overline{\scrP }_{h/A}^{in}\left( x,b,\{m_{q}\},\frac{C_{1}}{C_{2}}\right) =\int _{x}^{1}\frac{d\xi _{a}}{\xi _{a}}\nonumber \\
 &  & \times {\mathcal{C}}_{h/a}^{in}\left( \xhat ,\mu _{F}b,bM;\frac{C_{1}}{C_{2}}\right) f_{a/A}\left( \xi _{a},\left\{ \frac{\mu _{F}}{m_{q}}\right\} \right) .\nonumber \\
 &  & \label{PbFactorizationHeavy2} 
\end{eqnarray}

To perform this calculation, we consider a more elementary form of Eq.~(\ref{PbFactorizationHeavy2}),
which represents the leading regions in Feynman graphs in the limit $ Q\rightarrow \infty  $.
This elementary form can be found in Ref.~\cite{Collins:1982va}, where
it was derived in the case of $ e^{+}e^{-} $ hadroproduction. The function
$ \overline{\scrP }_{j/A}^{in}\left( x,b,\{m_{q}\},C_{1}/C_{2}\right)  $
is decomposed as
\begin{eqnarray}
 &  & \overline{\scrP }_{j/A}^{in}\left( x,b,\{m_{q}\},\frac{C_{1}}{C_{2}}\right) =\left| \scrH _{j}\left( \frac{C_{1}}{C_{2}b}\right) \right| \nonumber \\
 &  & \times \widetilde{U}\left( b\right) ^{1/2}\, \widehat{\scrP }_{j/A}^{in}\left( x,b,\{m_{q}\},\mu ,\frac{C_{1}}{C_{2}}\right) .\label{barPin} 
\end{eqnarray}
Here $ \scrH _{j} $ denotes the {}``hard vertex{}'', which contains
highly off-shell subgraphs. $ \widetilde{U} $ denotes soft subgraphs attached
to $ \scrH _{j} $ through gluon lines. $ \wh{\scrP }_{j/A}^{in}(x,b,\{m_{q}\},C_{1}/C_{2}) $
consists of subgraphs corresponding to the propagation of the incoming hadronic
jet. The jet part $ \wh{\scrP }_{j/A}^{in}(x,b,\{m_{q}\},C_{1}/C_{2}) $
is related to the $ k_{T} $-dependent PDF $ \scrP _{j/A}^{in}(x,k_{T},\{m_{q}\},\zeta _{A}) $,
defined as
\begin{eqnarray}
 &  & \scrP _{j/A}^{in}(x,k_{T},\{m_{q}\},\zeta _{A})=\overline{\sum _{spin}}\, \overline{\sum _{color}}\int \frac{dy^{-}d^{2}\vec{y}_{T}}{(2\pi )^{3}}\nonumber \\
 &  & \times e^{-ixp_{A}^{+}y^{-}+i\vec{k}_{T}\cdot \vec{y}_{T}}\nonumber \\
 &  & \times \langle p_{A}|\bar{\psi }_{j}(0,y^{-},\vec{y}_{T})\frac{\gamma ^{+}}{2}\psi _{j}(0)|p_{A}\rangle \label{Pin} 
\end{eqnarray}
in the frame where $ p_{A}^{\mu }=\left\{ p_{A}^{+},0,\vec{0}_{T}\right\} , $
$ p_{a}^{\mu }=\left\{ xp_{A}^{+},M^{2}/(2xp_{A}^{+}),\vec{k}_{T}\right\} , $
and $ p_{A}^{+}\rightarrow \infty  $. This definition is given in a gauge
$ \eta \cdot {\mathscr A}=0 $ with $ \eta ^{2}<0 $. The $ k_{T} $-dependent
PDF depends on the gauge through the parameter $ \zeta _{A}\equiv (p_{A}\cdot \eta )/|\eta ^{2}| $. 

Let $ \wt{\scrP }_{j/A}^{in}(x,b,\{m_{q}\},\zeta _{A}) $ be the $ b $-space
transform of $ \scrP _{j/A}^{in}(x,k_{T},\{m_{q}\},\zeta _{A}) $ taken
in $ d $ dimensions:
\begin{eqnarray}
 &  & \wt{\scrP }_{j/A}^{in}(x,b,\zeta _{A},\{m_{q}\})\equiv \int d^{d-2}\vec{k}_{T}e^{i\vec{k}_{T}\cdot \vec{b}}\nonumber \\
 &  & \times \scrP _{j/A}^{in}(x,k_{T},\zeta _{A},\{m_{q}\}).\label{wtPin} 
\end{eqnarray}
Note that our definition $ \wt{\scrP }_{j/A}^{in}(x,b,\zeta _{A},\{m_{q}\}) $
differs from the definition in Ref.~\cite{Collins:1982uw} by a factor $ (2\pi )^{2-d}. $
The jet part $ \wh{\scrP }_{a/A}^{in}(x,b,\{m_{q}\},C_{1}/C_{2}) $ is
related to $ \wt{\scrP }_{j/A}^{in}(x,b,\{m_{q}\},\zeta _{A}) $ in the
limit $ \zeta _{A}\rightarrow \infty  $: 
\begin{eqnarray}
 &  & \widehat{\scrP }_{j/A}^{in}\left( x,b,\{m_{q}\},\frac{C_{1}}{C_{2}}\right) =\lim _{\zeta _{A}\rightarrow \infty }\Biggl \{e^{{\cal S}'(b,\zeta _{A};C_{1}/C_{2})}\nonumber \\
 &  & \times \wt{\scrP }_{j/A}^{in}(x,b,\{m_{q}\},\zeta _{A})\Biggr \},\label{limzeta} 
\end{eqnarray}
where $ {\cal S}'(b,\zeta _{A};C_{1}/C_{2}) $ is a partial Sudakov factor,
\begin{eqnarray}
 &  & {\cal S}'(b,\zeta _{A};C_{1}/C_{2})\equiv \int _{C_{1}/b}^{C_{2}\zeta _{A}^{1/2}}\frac{d\bar{\mu }}{\bar{\mu }}\nonumber \\
 &  & \times \Biggl [\ln \left( \frac{C_{2}\zeta ^{1/2}}{\bar{\mu }}\right) \gamma _{\mathscr K}(\alpha _{S}(\bar{\mu }))\nonumber \\
 &  & -{\mathscr K}\left( b;\alpha _{S}\left( \frac{C_{1}}{b}\right) ,\frac{C_{1}}{b}\right) -{\mathscr G}\left( \frac{\bar{\mu }}{C_{2}};\alpha _{S}(\bar{\mu }),\bar{\mu }\right) \Biggr ].\nonumber \\
 &  & \label{Sprime} 
\end{eqnarray}
The definitions of the functions $ \gamma _{\mathscr K}, $ $ {\mathscr K} $,
and $ \mathscr G $ can be found in Ref.~\cite{Collins:1981uk}. 

We now have all necessary ingredients for the calculation of the $ {\cal O}(\alpha _{S}/\pi ) $
function $ {\cal C}^{in(1)}_{h/G}(x,\mu _{F}b,bM) $. Setting $ j=h $
and $ A=G, $ and expanding Eqs.~(\ref{PbFactorizationHeavy2},\ref{barPin},\ref{limzeta}),
and (\ref{Sprime}) in powers of $ \alpha _{S}/\pi  $, we find
\begin{eqnarray}
{\cal C}^{in(1)}_{h/G}(x,\mu _{F}b,bM) & = & \lim _{\zeta _{A}\rightarrow \infty }\left\{ \wt{\scrP }_{h/G}^{in(1)}(x,b,M,\zeta _{A})\right\} \nonumber \\
 & - & f_{h/G}^{(1)}\left( x,\mu _{F}/M\right) ,\label{Cin1hG} 
\end{eqnarray}
 where the superscript in parentheses denotes the order of $ \alpha _{S}/\pi  $.
In the derivation of this equation, we used the following easily deducible
equalities:
\begin{eqnarray}
 &  & \mathscr {H}_{h}^{(0)}=\wt{U}^{(0)}=1,\\
 &  & \left( {\cal S}^{'}\right) ^{(0)}=\wt{\scrP }_{h/G}^{in(0)}={\cal C}_{h/G}^{in(0)}=f_{h/G}^{(0)}=0,\\
 &  & {\cal C}_{h/h}^{in(0)}(x)=f_{G/G}^{(0)}(x)=\delta (x-1).
\end{eqnarray}

\begin{figure}
{\centering \resizebox*{!}{3cm}{\includegraphics{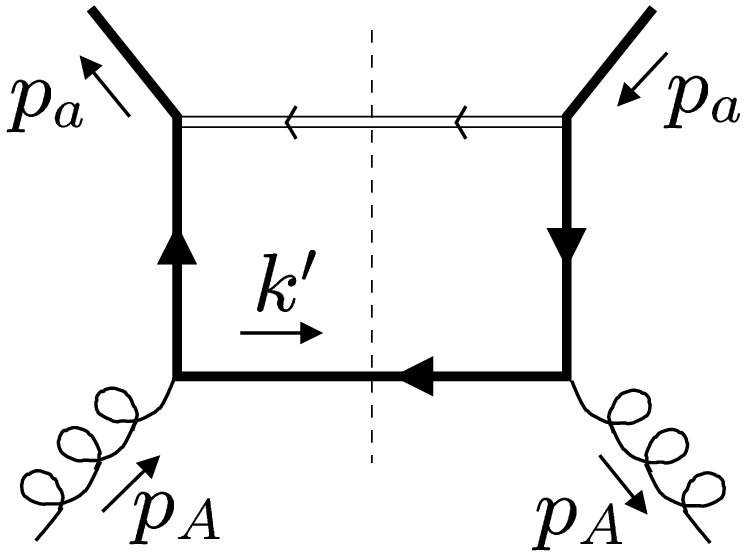}} \par}

\caption{\label{fig:pdf_diag}The Feynman diagram for the \protect$ {\cal O}(\alpha _{S}/\pi )\protect $
contributions \protect$ f_{h/G}^{(1)}\left( x,\mu _{F}/M\right) \protect $
and \protect$ \lim _{\zeta _{A}\rightarrow \infty }\wt{\scrP }_{h/G}^{in(1)}(x,b,M,\zeta _{A})\protect $.}
\end{figure}

The r.h.s.~of Eq.~(\ref{Cin1hG}) can be calculated with the help of the
definitions for $ f_{h/G}^{(1)}\left( x,\mu _{F}/M\right)  $ and $ \wt{\scrP }_{h/G}^{in(1)}(x,b,M,\zeta _{A}) $
in Eqs.~(\ref{f}) and (\ref{Pin},\ref{wtPin}), respectively. A further
simplification can be achieved by observing that at $ {\cal O}(\alpha _{S}/\pi ) $
the limit $ \eta ^{2}\rightarrow 0 $ in $ \wt{\scrP }_{h/G}^{in(1)}(x,b,M,\zeta _{A}) $
can be safely taken before the limit $ \zeta _{A}\rightarrow \infty  $,
and, furthermore, for $ \eta ^{2}=0 $ the function $ \wt{\scrP }_{h/G}^{in(1)}(x,b,M,\zeta _{A}) $
does not depend on $ \zeta _{A} $. Correspondingly, both objects can be
derived in the lightlike gauge from a single cut diagram shown in Fig.~\ref{fig:pdf_diag},
where the double line corresponds to the factor $ \gamma ^{+}\delta (p_{A}^{_{+}}-p^{+}_{a}-k^{'+})/2 $
in the case of $ f_{h/G}^{(1)}\left( x,\mu _{F}/M\right)  $ and $ \gamma ^{+}\delta (p_{A}^{_{+}}-p^{+}_{a}-k^{'+})e^{i\vec{k}_{T}^{'}\cdot \vec{b}}/2 $
in the case of $ \lim _{\zeta _{A}\rightarrow \infty }\wt{\scrP }_{h/G}^{in(1)}(x,b,M,\zeta _{A}) $. 

The difference between $ \lim _{\zeta _{A}\rightarrow \infty }\wt{\scrP }_{h/G}^{in(1)}(x,b,M,\zeta _{A})\equiv \wt{\scrP }_{h/G}^{in(1)}(x,b,M) $
and $ f_{h/G}^{(1)}\left( x,\mu _{F}/M\right)  $ resides in the extra
exponential factor $ e^{i\vec{k}_{T}^{'}\cdot \vec{b}} $ in $ \wt{\scrP }_{h/G}^{in(1)}(x,b,M) $.
Remarkably, this factor strongly affects the nature of $ \wt{\scrP }_{h/G}^{in(1)}(x,b,M) $.
The loop integral over $ \vec{k}_{T}^{'} $ in $ f_{h/G}^{(1)}\left( x,\mu _{F}/M\right)  $
contains a UV singularity, which is regularized by an appropriate counterterm.
In the ACOT scheme, the UV singularity is regularized in the $ \overline{MS} $
scheme if $ \mu _{F}\geq M $, and by zero-momentum subtraction if $ \mu _{F}<M $.
The result for the heavy-quark PDF $ f_{h/G}^{(1)}\left( x,\mu _{F}/M\right)  $
is
\begin{eqnarray}
f^{(1)}_{h/G}\left( x,\frac{\mu _{F}}{M}\right)  & = & \left\{ \begin{array}{c}
P_{h/G}^{(1)}(x)\ln \left( \mu _{F}/M\right) ,\, \mu _{F}\geq M;\\
0,\, \mu _{F}<M.
\end{array}\right. \nonumber \\
 &  & \label{f1} 
\end{eqnarray}
 As expected, $ f_{h/G}^{(1)}\left( x,\mu _{F}/M\right)  $ exhibits the
threshold behavior at $ \mu _{F}=M $.

In contrast, the UV limit in the loop integral of $ \wt{\scrP }_{h/G}^{in(1)}(x,b,M) $
is regularized by the oscillating exponent $ e^{i\vec{k}_{T}^{'}\cdot \vec{b}} $.
Since no UV singularity is present in $ \wt{\scrP }_{h/G}^{in(1)}(x,b,M) $,
it does not depend on $ \mu _{F} $ and, therefore, does not change at
the threshold. It is given by
\begin{eqnarray}
\wt{\scrP }_{h/G}^{in(1)}(x,b,M) & = & P^{(1)}_{h/G}(x)K_{0}(bM)\nonumber \\
 & + & T_{R}x(1-x)bMK_{1}(bM).\label{PinhG1} 
\end{eqnarray}
Here $ K_{0}(bM) $ and $ K_{1}(bM) $ are the modified Bessel functions
\cite{AbramowitzStegun}, which satisfy the following useful properties:
\begin{eqnarray}
 &  & \lim _{bM\rightarrow \infty }K_{0}(bM)=\lim _{bM\rightarrow \infty }bMK_{1}(bM)=0;\label{K0infty} \\
 &  & K_{0}(bM)\rightarrow -\ln \left( bM/b_{0}\right) \mbox {\, as\, }bM\rightarrow 0;\\
 &  & bM\, K_{1}(bM)\rightarrow 1\mbox {\, as\, }bM\rightarrow 0.\label{K10} 
\end{eqnarray}
The {}``infrared-safe{}'' part $ {\mathcal{C}}^{in(1)}_{h/G}(x,\mu _{F}b,bM) $
of $ \wt{\scrP }_{h/G}^{in(1)}(x,b,M) $ is obtained by subtracting $ f^{(1)}_{h/G}\left( x,\mu _{F}/M\right)  $
as in Eq.~(\ref{Cin1hG}):
\begin{eqnarray}
 &  & \left. {\mathcal{C}}^{in(1)}_{h/G}(x,\mu _{F}b,bM)\right| _{\mu _{F}\geq M}=T_{R}x(1-x)\nonumber \\
 &  & \times \left( 1+c_{1}(bM)\right) \nonumber \\
 &  & +P^{(1)}_{h/G}(x)\left( c_{0}(bM)-\ln \Bigl (\frac{\mu _{F}b}{b_{0}}\Bigr )\right) ;\\
 &  & \left. {\mathcal{C}}^{in(1)}_{h/G}(x,\mu _{F}b,bM)\right| _{\mu _{F}<M}=\wt{\scrP }_{h/G}^{in(1)}(x,b,M)\nonumber \\
 &  & =\left. {\mathcal{C}}^{in(1)}_{h/G}(x,\mu _{F}b,bM)\right| _{\mu _{F}\geq M}\nonumber \\
 &  & +P^{(1)}_{h/G}(x)\ln \frac{\mu _{F}}{M}.\label{C1injG} 
\end{eqnarray}
 In these equations, $ c_{0}(bM) $ and $ c_{1}(bM) $ are the parts
of $ K_{0}(bM) $ and $ bM\, K_{1}(bM) $ that vanish at $ bM\rightarrow 0 $
(cf. Eqs.~( \ref{K0infty}-\ref{K10})):
\begin{eqnarray}
c_{0}(bM) & \equiv  & K_{0}(bM)+\ln \frac{bM}{b_{0}};\label{c0} \\
c_{1}(bM) & \equiv  & bMK_{1}(bM)-1.\label{c1} 
\end{eqnarray}
 If $ \mu _{F} $ is chosen to be of order $ b_{0}/b $, no large logarithms
appear in $ {\mathcal{C}}^{in(1)}_{h/G}(x,\mu _{F}b,bM) $ at $ b\rightarrow 0 $.
At large $ Q $, the small-$ b $ region dominates the integration in
Eq.~(\ref{Wmassive}), so that $ {\mathcal{C}}^{in}_{h/G}(\xhat ,\mu _{F}b,bM) $
effectively reduces to its massless expression \cite{Meng:1996yn, Nadolsky:1999kb}:
\begin{eqnarray}
 &  & \left. {\mathcal{C}}^{in(1)}_{h/G}(x,\mu _{F}b,bM)\right| _{b\rightarrow 0}=T_{R}x(1-x)\nonumber \\
 &  & -P^{(1)}_{h/G}(x)\ln \Bigl (\frac{\mu _{F}b}{b_{0}}\Bigr ).
\end{eqnarray}

The above manipulations can be interpreted in the following way. At small
$ b $ ($ b=b_{0}/\mu _{F}\leq b_{0}/M $), we subtract from $ \wt{\scrP }_{h/G}^{in(1)}(x,b,M) $
its infrared-divergent part $ P^{(1)}_{h/G}(x)\ln (\mu _{F}/M), $ which
is then included and resummed in the heavy-quark PDF $ f_{h/G}(x,\mu _{F}/M) $.
The convolution of the resulting $ {\cal C} $-function with the PDF remains
equal to $ \wt{\scrP }_{h/G}^{in(1)}(x,b,M) $ up to higher-order corrections:
\begin{eqnarray}
 &  & \sum _{a=h,G}{\cal C}^{in}_{h/a}\otimes f_{a/G}=\overline{\scrP }^{in(1)}_{h/G}(x,\mu _{F}b,bM)+{\mathcal{O}}(\alpha _{S}^{2}).\nonumber \\
 &  & \label{cfs} 
\end{eqnarray}
At large $ b $ ($ b>b_{0}/M $), the heavy-quark PDF $ f_{h/G} $
is identically equal to zero. To preserve the relationship (\ref{cfs}) below
the threshold, one should include the above logarithmic term in the function
$ {\mathcal{C}}^{in(1)}_{h/G}(x,\mu _{F}b,bM) $, as shown in Eq.~(\ref{C1injG}).
The addition of an extra term $ P^{(1)}_{h/G}(x)\ln \left( \mu _{F}/M\right)  $
to $ {\mathcal{C}}^{in(1)}_{h/G}(x,\mu _{F}b,bM) $ at $ \mu _{F}<M $
enforces the smoothness of the form-factor $ \wt{W}(b,Q,x,z) $ in the
threshold region, which, in its turn, is needed to avoid unphysical oscillations
of the cross section $ d\sigma /dq_{T}^{2} $.

\section{\label{Appendix:FO}The finite-order cross section }

This Appendix discusses the finite-order cross section $ d\wh{\sigma }_{FO}/d\wh{\Phi } $
that appears in the factorized hadronic cross section (\ref{Factorization}).
For the $ {\cal O}(\alpha _{S}^{0}) $ subprocess $ e+h\rightarrow e+h $,
this cross section is the same as in the massless case:
\begin{eqnarray}
 &  & \left( \frac{d\wh{\sigma }(e+h\rightarrow e+h)}{d\wh{\Phi }}\right) _{FO}=\frac{\sigma _{0}F_{l}}{S_{eA}}\frac{A_{1}(\psi ,\varphi )}{2}\nonumber \\
 &  & \times e_{j}^{2}\delta (\vec{q}_{T})\delta (1-\xhat )\delta (1-\zhat ),\label{FiniteOrderLO} 
\end{eqnarray}
where, in accordance with the notations of Ref.~\cite{Nadolsky:1999kb},
\begin{eqnarray}
\sigma _{0} & \equiv  & \frac{Q^{2}}{4\pi S_{eA}x^{2}}\Bigl (\frac{e^{2}}{2}\Bigr ),\nonumber \label{sigma0Fl} \\
F_{l} & \equiv  & \frac{e^{2}}{2}\frac{1}{Q^{2}}.
\end{eqnarray}

The contribution of the gluon-photon fusion channel is 
\begin{widetext}
\begin{eqnarray}
 &  & \left( \frac{d\wh{\sigma }(e+G\rightarrow e+h+\bar{h})}{d\wh{\Phi }}\right) _{FO}=\frac{\sigma _{0}F_{l}}{4\pi S_{eA}}\frac{\alpha _{S}}{\pi }e_{Q}^{2}\delta \left( \left( \frac{1}{\xhat }-1\right) \left( \frac{1}{\zhat }-1\right) -\frac{\wt{q}_{T}^{2}}{Q^{2}}\right) \frac{\xhat (1-\xhat )}{\zhat ^{2}}\nonumber \\
 &  & \qquad \qquad \times T_{R}\sum _{\rho =1}^{4}\wh{V}_{\rho }(\xhat ,Q^{2},\zhat ,q_{T}^{2},M^{2})A_{\rho }(\psi ,\varphi ),\label{FiniteOrder} 
\end{eqnarray}
\end{widetext} where $ A_{\rho }(\psi ,\varphi ) $ denote orthonormal
functions of the leptonic azimuthal angle $ \varphi  $ and boost parameter
$ \psi  $ given in Eq.~(\ref{As}). The structure functions $ \wh{V}_{\rho }(\xhat ,Q^{2},\zhat ,q_{T}^{2},M^{2}) $
are calculated to be
\begin{eqnarray}
\wh{V}_{1} & = & \frac{1}{\xhat ^{2}\wt{q}_{T}^{2}}\left( 1-2\xhat \zhat +2\xhat ^{2}\zhat ^{2}-4\frac{M^{2}\xhat ^{2}}{Q^{2}}\right) \nonumber \\
 & + & \frac{2\zhat }{\xhat }\frac{1}{Q^{2}}\left( 5\xhat \zhat -\xhat -\zhat \right) +\nonumber \\
 & + & \kappa _{1}\left( 4\frac{\wt{q}_{T}^{2}}{Q^{2}}\zhat ^{2}+2-8\zhat +8\zhat ^{2}-4\frac{M^{2}}{Q^{2}}\right) ,\label{V1jg} \\
\wh{V}_{2} & = & 8\frac{1}{Q^{2}}\zhat ^{2}-4\frac{M^{2}}{Q^{2}}\frac{1}{\wt{q}_{T}^{2}}+4\kappa _{1}\left( -1+\zhat \right) \zhat ,\\
\wh{V}_{3} & = & \frac{2\zhat }{\xhat }\frac{q_{T}}{Q\wt{q}_{T}^{2}}\left( -1+2\left( 1+\frac{\wt{q}_{T}^{2}}{Q^{2}}\right) \xhat \zhat \right) \nonumber \\
 & + & 4\kappa _{1}\zhat \left( -1+2\zhat \right) \frac{q_{T}}{Q},\\
\wh{V}_{4} & = & 4\frac{q_{T}^{2}}{Q^{2}\wt{q}_{T}^{2}}\zhat ^{2}+4\frac{q_{T}^{2}}{Q^{2}}\zhat ^{2}\kappa _{1}.\label{V4jg} 
\end{eqnarray}
 In Eqs.~(\ref{V1jg}-\ref{V4jg}),
\begin{equation}
\kappa _{1}\equiv \frac{M^{2}(1-\xhat )}{\zhat ^{2}\xhat \wt{q}_{T}^{4}}.
\end{equation}

\section{\label{Appendix:KinematicalCorrection}Kinematical correction}

In this Appendix, we derive the kinematical corrections (\ref{chih1}) and
(\ref{chih2}) that are introduced in the flavor-excitation contributions
to $ \sigma _{FO} $, as well as in $ \sigma _{\widetilde{W}} $ and
$ \sigma _{ASY} $. Let us first consider the $ {\cal O}(\alpha _{S}) $
cross section for the photon-gluon fusion, which we write as
\begin{eqnarray}
 &  & \left( \frac{d\sigma (e+A\rightarrow e+H+X)}{d\Phi }\right) _{\gamma ^{*}G,FO}=\nonumber \\
 &  & \int \frac{d\xi _{b}}{\xi _{b}}\int \frac{d\xi _{a}}{\xi _{a}}D_{H/h}\left( \xi _{b}\right) f_{G/A}\left( \xi _{a}\right) \nonumber \\
 &  & \times \delta \left( \left( \frac{1}{\xhat }-1\right) \left( \frac{1}{\zhat }-1\right) -\frac{\wt{q}_{T}^{2}}{Q^{2}}\right) \beta (\Phi ).\label{FiniteOrder2} 
\end{eqnarray}
Here $ \beta (\wh{\Phi }) $ includes all terms in the parton-level cross
section $ (d\wh{\sigma }/d\wh{\Phi })_{FO} $ except for the $ \delta - $function
(cf. Eq.~(\ref{FiniteOrder})):
\begin{eqnarray}
 &  & \beta (\wh{\Phi })=\frac{\sigma _{0}F_{l}}{4\pi S_{eA}}\frac{\alpha _{S}}{\pi }e_{h}^{2}\frac{\xhat (1-\xhat )}{\zhat ^{2}}\nonumber \\
 & \times  & T_{R}\sum _{\rho =1}^{4}\wh{V}_{\rho }(\xhat ,Q^{2},\zhat ,q_{T}^{2},M^{2})A_{\rho }(\psi ,\varphi ).
\end{eqnarray}

The $ \delta  $-function in Eq.~(\ref{FiniteOrder2}) can be reorganized
as
\begin{eqnarray}
 &  & \delta \left( \left( \frac{1}{\xhat }-1\right) \left( \frac{1}{\zhat }-1\right) -\frac{\wt{q}_{T}^{2}}{Q^{2}}\right) =\nonumber \\
 &  & \frac{z\, Q^{2}}{\sqrt{\widehat{W}^{4}-4M^{2}\left( q_{T}^{2}+\widehat{W}^{2}\right) }}\Biggl [\delta \left( \xi _{b}-\xi _{b}^{+}\right) \nonumber \\
 &  & +\delta \left( \xi _{b}-\xi _{b}^{-}\right) \Biggr ],\label{deltaxib} 
\end{eqnarray}
 where
\begin{equation}
\xi _{b}^{\pm }\equiv z\frac{\widehat{W}^{2}\pm \sqrt{\widehat{W}^{4}-4M^{2}(q_{T}^{2}+\widehat{W}^{2})}}{2M^{2}},
\end{equation}
and $ \wh{W}^{2}\equiv Q^{2}\left( 1-\xhat \right) /\xhat  $. We see that
the mass-dependent phase space element contains two $ \delta  $-functions
$ \delta (\xi _{b}-\xi _{b}^{+}) $ and $ \delta (\xi _{b}-\xi _{b}^{-}) $,
which can be used to integrate out the dependence on $ \xi _{b} $ in Eq.~(\ref{FiniteOrder2}). 

It can be further shown that in the massless limit the solutions $ \xi _{b}=\xi _{b}^{-} $
and $ \xi _{b}=\xi _{b}^{+} $ correspond to the heavy quarks produced
in the current and target fragmentation regions, respectively. When $ M\rightarrow 0 $,
the relationship (\ref{deltaxib}) simplifies to
\begin{eqnarray}
 &  & \delta \left( \left( \frac{1}{\xhat }-1\right) \left( \frac{1}{\zhat }-1\right) -\frac{q_{T}^{2}}{Q^{2}}\right) =\nonumber \\
 &  & \frac{z\, Q^{2}}{\widehat{W}^{2}}\left[ \delta \left( \xi _{b}-\xi _{b}^{0+}\right) +\delta \left( \xi _{b}-\xi _{b}^{0-}\right) \right] ,\label{deltaxibM0} 
\end{eqnarray}
 where
\begin{eqnarray}
\xi _{b}^{0+} & = & z\left( \frac{\widehat{W}^{2}}{M^{2}}-\frac{q_{T}^{2}+\widehat{W}^{2}}{\wh{W}^{2}}+{\mathcal{O}}(M^{2})\right) ,\\
\xi _{b}^{0-} & = & z\left( \frac{q_{T}^{2}+\widehat{W}^{2}}{\widehat{W}^{2}}+{\mathcal{O}}(M^{2})\right) .
\end{eqnarray}
In this limit, the solution $ \xi _{b}^{0+} $ diverges (and, therefore,
will not contribute) unless $ z $ is identically zero. However, according
to Eq.~(\ref{z}) and the last paragraph in Subsection B of Section~\ref{sec:PhotonGluon},
at $ z=0 $ the observed final-state hadron appears among remnants of the
target ($ \theta _{H}\approx 180^{\circ } $ in the $ \gamma ^{*}A $
c.m.~frame), \emph{i.e.}, \emph{away} from the region of our primary interest
(small and intermediate $ \theta _{H} $). Hence, in the limit $ \theta _{H}\rightarrow 0 $
all dominant logarithmic contributions as well as their all-order sums (the
flavor-excitation cross section and $ \widetilde{W} $-term) arise only
from terms proportional to $ \delta (\xi _{b}-\xi _{b}^{-}) $. The contributions
proportional to $ \delta (\xi _{b}-\xi _{b}^{+}) $ in the current fragmentation
region are suppressed. 

The integration over $ \xi _{b} $ with the help of Eq.~(\ref{deltaxib})
leads to the following expression for the cross section (\ref{FiniteOrder2}):
\begin{widetext}
\begin{eqnarray}
 &  & \left( \frac{d\sigma (e+A\rightarrow e+H+X)}{dxdQ^{2}dzdq_{T}^{2}d\varphi }\right) _{\gamma ^{*}G}=\int _{\xi ^{min}_{a}}^{\xi _{a}^{max}}\frac{d\xi _{a}}{\xi _{a}}f_{G/A}(\xi _{a},\mu _{F})\frac{Q^{2}}{\sqrt{\widehat{W}^{4}-4M^{2}\left( q_{T}^{2}+\widehat{W}^{2}\right) }}\nonumber \\
 &  & \times \left[ \left. \zhat D_{H/h}(\xi _{b},\mu _{F})\beta (\wh{\Phi })\right| _{\xi _{b}=\xi _{b}^{+}}+\left. \zhat D_{H/h}(\xi _{b},\mu _{F})\beta (\wh{\Phi })\right| _{\xi _{b}=\xi _{b}^{-}}\right] .\label{Fobeta2} 
\end{eqnarray}
\end{widetext} Here the lower and upper integration limits $ \xi _{a}^{min} $
and $ \xi _{a}^{max} $ are determined by demanding the argument of the
square root in Eq.~(\ref{Fobeta2}) be non-negative and $ \xi _{b}\leq 1 $;
that is,
\begin{eqnarray}
\xi _{a}^{min} & = & x\left( 1+\frac{2M\left( M+\sqrt{M^{2}+q_{T}^{2}}\right) }{Q^{2}}\right) ,\nonumber \\
\xi ^{max}_{a} & = & \min \left[ x\left( 1+\frac{M^{2}+z^{2}q_{T}^{2}}{z(1-z)Q^{2}}\right) ,1\right] \label{ximinximaxPlus} 
\end{eqnarray}
for $ \xi _{b}=\xi _{b}^{+} $, and
\begin{eqnarray}
\xi _{a}^{min} & = & x\left( 1+\frac{1}{z(1-z)}\frac{M^{2}+z^{2}q_{T}^{2}}{Q^{2}}\right) ,\nonumber \\
\xi ^{max}_{a} & = & 1\label{ximinximaxMinus} 
\end{eqnarray}
for $ \xi _{b}=\xi _{b}^{-} $. We see that, according to the exact kinematics
of heavy flavor production, the heavy quark pairs are produced only when
the light-cone momentum fraction $ \xi _{a} $ is not less than $ \xi _{a}^{min} $
(where $ \xi _{a}^{min}\geq x $) and not more than $ \xi _{a}^{max} $
(where $ \xi _{a}^{max}\leq 1 $). The exact values of $ \xi _{a}^{min} $
and $ \xi _{a}^{max} $ are different for the branches with $ \xi _{b}=\xi _{b}^{+} $
and $ \xi _{b}=\xi _{b}^{-}. $ 

Turning now to the flavor-excitation contributions $ \gamma ^{*}+h\rightarrow h+X $,
we find that in those the integration over $ \xi _{a} $ \emph{a priori}
covers the whole range $ x\leq \xi _{a}\leq 1. $ Indeed, in those contributions
the heavy antiquark in the remnants of the incident hadron is ignored, so
that the reaction can go at a lower c.m.~energy $ \widehat{W} $ than
it is allowed by the exact kinematics. Since the PDF's grow rapidly at small
$ x $, the naively calculated total cross section $ \sigma _{TOT} $
tends to contain large contributions from the unphysical region of small
$ x $ and disagree with the data. To fix this problem, we use Eq.~(\ref{ximinximaxMinus})
to derive the following scaling variable in the \emph{finite-order} flavor-excitation
contributions:
\begin{equation}
\label{chi21}
\chi _{h}=x\left( 1+\frac{1}{z(1-z)}\frac{M^{2}}{Q^{2}}\right) .
\end{equation}
 This variable takes into account the fact that the incoming heavy quark
in the flavor-excitation process appears from the contributions with $ \xi _{b}=\xi _{b}^{-} $
in the flavor-creation process, and that the transverse momentum $ zq_{T} $
of this quark in the \emph{finite-order} cross section is identically zero.

Similarly, we notice that the $ \widetilde{W} $-term $ \sigma _{\widetilde{W}} $
and its finite-order expansion $ \sigma _{ASY} $ contain the {}``$ b $-dependent
PDF's{}'' $ \overline{\scrP }_{h/A}^{in}\left( x,b,M,C_{1}/C_{2}\right)  $,
which correspond to the incoming heavy quarks with a \emph{non-zero} transverse
momentum. According to Eq.~(\ref{ximinximaxMinus}), the available phase
space in the longitudinal direction is a decreasing function of the transverse
momentum $ zq_{T} $, and it is desirable to implement this phase-space
reduction to improve the cancellation between $ \sigma _{\widetilde{W}} $
and $ \sigma _{ASY} $ at large $ q_{T} $. In our calculation, this
feature is implemented by evaluating $ \sigma _{\widetilde{W}} $ and $ \sigma _{ASY} $
at the scaling variable
\begin{equation}
\label{chi22}
\chi '_{h}=x\left( 1+\frac{1}{z(1-z)}\frac{M^{2}+z^{2}q_{T}^{2}}{Q^{2}}\right) ,
\end{equation}
 which immediately follows from Eq.~(\ref{ximinximaxMinus}). 

Despite the apparent complexity of the scaling variables (\ref{chi21}) and
(\ref{chi22}), they satisfy the following important properties:

\begin{enumerate}
\item They are straightforwardly derived from the exact kinematical constraints
on the variable $ \xi _{a} $ in Eqs.~(\ref{ximinximaxPlus}) and (\ref{ximinximaxMinus}).
\item They remove contributions from unphysical values of $ x $ \emph{at all
values of $ Q $ and $ q_{T} $}, thus leading to better agreement
with the data.
\item In the limit $ Q^{2}\gg M^{2}, $ the variable $ \chi _{h} $ in $ \sigma _{FO} $
reduces to $ x $ (cf.~Eq.~(\ref{chi21})), so that the standard factorization
for the massless finite-order cross sections is reproduced. 
\item In the limit $ Q^{2}\gg M^{2} $ and $ Q^{2}\gg q_{T}^{2} $, the variable
$ \chi _{h}^{'} $ in $ \sigma _{\widetilde{W}} $ and $ \sigma _{ASY} $
reduces to $ x $ (cf.~Eq.~(\ref{chi22})), so that the exact resummed
cross section is reproduced. 
\end{enumerate}
Finally, consider the integration of the cross section (\ref{Fobeta2}) over
$ z, $ $ q^{2}_{T} $, and $ \varphi  $ to obtain the $ {\cal O}(\alpha _{S}) $
$ \gamma ^{*}G $ contribution to an inclusive DIS function $ F(x,Q^{2}) $.
We find that
\begin{eqnarray}
\left. F(x,Q^{2})\right| _{\gamma ^{*}G,{\cal O}(\alpha _{S})} & = & \int _{\xi _{a}^{'}}^{1}\frac{d\xi _{a}}{\xi _{a}}C^{(1)}_{H/G}\left( \frac{x}{\xi _{a}},\frac{\mu _{F}}{Q},\frac{M}{Q}\right) \nonumber \\
 & \times  & f_{G/A}\left( \xi _{a},\, \left\{ \frac{\mu _{F}}{m_{q}}\right\} \right) ,
\end{eqnarray}
 where the lower limit of the integral over $ \xi _{a} $ is given by
\begin{equation}
\xi _{a}^{'}=x\left( 1+\frac{4M^{2}}{Q^{2}}\right) 
\end{equation}
 for both solutions $ \xi _{b}=\xi _{b}^{+} $ and $ \xi _{b}=\xi _{b}^{-} $.
This value of $ \xi _{a}^{'} $ can be easily found from Eqs.~(\ref{ximinximaxPlus})
and (\ref{ximinximaxMinus}), given that $ q_{T}\geq 0 $, $ 0\leq z\leq 1 $,
and $ z(1-z)\leq 1/4 $ in the interval $ 0\leq z\leq 1 $. Since in
the $ \gamma ^{*}G $ contribution the integration over $ \xi _{a} $
is constrained from below by $ \xi _{a}^{'}>x $, it makes sense to implement
a similar constraint in the flavor-excitation contributions by introducing
the scaling variable $ \chi _{h}=x(1+4M^{2}/Q^{2}). $ This variable is
precisely the one that appears in the recent version of the ACOT scheme with
the optimized treatment of the inclusive structure functions in the threshold
region \cite{Tung:2001mv}. Our scaling variables extend the idea of Ref.~\cite{Tung:2001mv}
to the semi-inclusive and resummed cross sections.

\newpage


\end{document}